%% file: main.tex
\documentclass[a4paper,UKenglish,cleveref, autoref, thm-restate]{lmcs}
\pdfoutput=1

\usepackage[utf8]{inputenc}
\usepackage{lastpage}
\lmcsdoi{18}{2}{23}
\lmcsheading{}{\pageref{LastPage}}{}{}%
{Jun.~16,~2021}{Jun.~29,~2022}{}

\usepackage{babel}
\usepackage{hyperref}
\usepackage{comment}

\newcommand{\Aa}{\mathcal{A}}
\newcommand{\Nn}{\mathbb{N}}
 
\newcommand{\Tt}{\mathcal{T}}

\newcommand{\Lang}{\mathcal{L}}
\newcommand\sem[1]{[\![ #1 ]\!]}
\usepackage{xspace}

\newcommand{\twoDFTla}{2\textsf{DMT}\textsubscript{\textsf{la}}\xspace}
\newcommand{\twoDFTpla}{2\textsf{DFT}\textsubscript{\textsf{pla}}\xspace}
\newcommand{\twoDBT}{2\textsf{DBT}\xspace}
\newcommand{\twoDFT}{2\textsf{DFT}\xspace}
\newcommand{\twoPA}{2\textsf{PA}\xspace}
\usepackage{amsmath} 
\newcommand\dom[1]{\mathsf{dom}(#1)}
\newcommand\pref[3]{\mathsf{Pref_{#1}}(#2, #3)}
\newcommand\out[1]{\mathsf{out}(#1)}

\newcommand{\leftend}{\vdash}
\newcommand{\upward}[1]{{\uparrow}#1}

\usepackage[ruled,linesnumbered]{algorithm2e}
\newcommand{\prefix}{\preceq}
\usepackage{tikz}
\usetikzlibrary{shapes}
\usetikzlibrary{trees}
\usetikzlibrary{patterns}
\newcommand\mismatch{\mathsf{mismatch}}

\newcommand{\Inf}{\mathsf{Inf}}

\newcommand{\ou}{$\mathsf{out}$}
\newcommand{\buchi}{B\"{u}chi}

\newcommand{\nlogspace}{\textsc{NLogSpace}}
\newcommand{\set}[1]{\left\{#1\right\}}
\newcommand{\tuple}[1]{\left(#1\right)}

\newcommand{\rat}[1]{\mathsf{Reg}(#1)}

\newcommand{\ie}{\textit{i.e.}~}
\newcommand{\eg}{\textit{e.g.}~}

\usetikzlibrary{automata, arrows}
\definecolor{gold}{rgb}{0.99,0.78,0.07}
\tikzstyle{background}=[rectangle,fill=gold, inner sep=0.08cm, rounded corners=1mm]

\newtheorem{claim}[thm]{Claim}

\keywords{transducers, infinite words, computability, continuity, synthesis}

\begin{document}

\title[Synthesis of Computable Regular Functions of Infinite Words]{Synthesis of Computable Regular Functions\texorpdfstring{\\}{ }of Infinite Words}

\author[V.~Dave]{Vrunda Dave}[a]	
\address{IIT Bombay, India}	
\email{\{vrunda,krishnas\}@cse.iitb.ac.in}  

\author[E.~Filiot]{Emmanuel Filiot}[b]	
\address{Universit\'{e} Libre de Bruxelles, Belgium}	
\email{efiliot@ulb.ac.be}  
\thanks{This work is partially supported by the MIS project F451019F (F.R.S.-FNRS) and the EOS project Verifying Learning Artificial Intelligence Systems (F.R.S.-FNRS and FWO). Emmanuel Filiot is a research associate at F.R.S.-FNRS}	

\author[S.~N.~Krishna]{Shankara Narayanan Krishna}[a]	

\author[N.~Lhote]{Nathan Lhote}[c]	
\address{Aix Marseille Univ, Universit\'{e} de Toulon, CNRS, LIS, Marseille, France}	
\email{nathan.lhote@lis-lab.fr}  
\thanks{This work is partially supported by the ERC Consolidator grant LIPA683080, as well as ANR DELTA (grant ANR-16-CE40-0007)}


\begin{abstract}
Regular functions from infinite words to infinite
words can be equivalently specified by MSO-transducers, streaming
$\omega$-string transducers as well as deterministic two-way
transducers with look-ahead. In their one-way restriction, the latter
transducers define the class of rational functions. Even though
regular functions are robustly characterised by several finite-state devices,
even the subclass of rational functions may contain functions which
are not computable (by a Turing machine with infinite input). 
This paper proposes a decision procedure for the following synthesis problem:
given a regular function $f$ (equivalently specified by one of the
aforementioned transducer model), is $f$ computable and if it is,
synthesize a Turing machine computing it. 

For regular functions, we show that computability is
equivalent to continuity, and therefore the problem boils down to
deciding continuity. We establish a generic
characterisation of continuity for functions
preserving regular languages under inverse image (such as regular
functions). We exploit this characterisation to show the decidability
of continuity (and hence computability) of rational and regular 
functions. For rational functions, 
we show that this can be done  in \textsc{NLogSpace} (it was already known to be in \textsc{PTime} by
Prieur). In a similar fashion, we also
effectively characterise uniform continuity of regular functions, and
relate it to the notion of uniform computability, which offers
stronger efficiency guarantees.  
\end{abstract}
\maketitle


\section{Introduction}\label{sec:intro}
Let $\textsf{Inputs}$ and $\textsf{Outputs}$ be two arbitrary sets of
elements called inputs and outputs respectively. 
A general formulation of the synthesis problem is as follows: for a given specification
of a function $S \colon \textsf{Inputs}\rightarrow 2^{\textsf{Outputs}}$ relating any input
$u\in\dom{S}$\footnote{$\dom{S}$ is the domain of $S$, \ie the
	set of inputs that have a non-empty image by $S$.} to a set of 
outputs $S(u)\subseteq
\textsf{Outputs}$, decide whether there exists a
(total) function $f \colon
\dom{S}\rightarrow \textsf{Outputs}$ such that $(i)$ for all
$u\in \dom{S}$, $f(u)\in S(u)$ and $(ii)$ $f$ satisfies some
additional constraints such as being computable in some way, by some
device which is effectively returned by the synthesis
procedure.
Assuming the axiom of choice, the relaxation of this
problem without constraint $(ii)$ always has a positive
answer. 
However with additional requirement $(ii)$, a function $f$ realising
$S$ may not exist in general. In this paper, we consider the particular case
where the specification $S$ is \emph{functional},\footnote{In this case, we just write $S(u)=v$
	instead of $S(u)=\{v\}$.} in the sense that $S(u)$ 
	is singleton set for all $u \in \dom{S}$.
Even in this particular case, $S$ may not be realisable while satisfying requirement $(ii)$.

The latter observation on the functional case can already be made in the Church
approach to synthesis~\cite{Chur62,BuLa69}, for which $\textsf{Inputs},\textsf{Outputs}$ are sets
of infinite words and $f$ is required to be implementable by a Mealy
machine (a deterministic automaton which can output symbols). More
precisely, an infinite word $\alpha$ over a finite alphabet $\Sigma$ is a function $\alpha \colon \mathbb{N} \rightarrow \Sigma$ and is written as 
$\alpha=\alpha(0)\alpha(1) \ldots$. The set of infinite words over
$\Sigma$ is denoted by $\Sigma^\omega$. In Church $\omega$-regular
synthesis, we have $\textsf{Inputs} = \Sigma_i^\omega$ and
$\textsf{Outputs} = \Sigma_o^\omega$, and  functions $S$ are specified by
$\omega$-automata over $\Sigma_i.\Sigma_o$. Thus, such an automaton defines
a language $L\subseteq (\Sigma_i.\Sigma_o)^\omega$ and in turn,
through projection, a function $S_L$ defined by $S_L(i_1i_2\dots) = \{ o_1o_2\dots \mid
i_1o_1i_2o_2\dots \in L\}$. Such a
specification is said to be synchronous, meaning that it alternatively reads an input symbol and produces an output symbol. 
It is also $\omega$-regular because
it can be represented as an automaton over $\Sigma_i.\Sigma_o$. 
As an example, consider $\Sigma_i =
\Sigma_o = \{ a,b,@\}$ and the function $S_{\sf swap}$ defined only for all
words of the form $u_1\sigma @u_2$ such that $u_1\in\{a,b\}^*$ and $\sigma\in\{a,b\}$ by 
$S_{\sf swap}(u_1\sigma @ u_2) = \sigma u_1 @ u_2$. The specification $S_{\sf swap}$ is
easily seen to be synchronous and $\omega$-regular, but not realisable
by any Mealy machine. This is because a Mealy machine is an input deterministic model and so cannot guess the last symbol before the $@$ symbol.

\subsection{Computability of functions over infinite words}

In Church synthesis, the notion of computability used for requirement
$(ii)$ is that of being computable by a Mealy machine. While this
makes sense in a reactive scenario where output symbols (reactions)
have to be produced immediately after input symbols are received, this
computability notion is too strong in a more relaxed scenario where
reactivity is not required. Instead, we propose here to investigate
the synthesis problem for functional specifications over infinite words
where the computability assumption $(ii)$ for $f$ is
just being computable by some algorithm (formally a Turing
machine) running on infinite inputs. In other words, our goal is to 
synthesize \emph{algorithms} from specifications of functions of
infinite words. There are classical
computability notions for infinite objects, like infinite sequences 
of natural numbers, motivated by real analysis, or computation of
functions of real numbers. The model of computation we consider for infinite words is a deterministic machine with three tapes: a read-only 
one-way tape holding the input, a two-way working tape  
with no restrictions and a write-only one-way output tape. All three
tapes are infinite on the right.
A function $f$ is computable if there
exists such a machine $M$ such that, if its input tape is fed with an
infinite word $x$ in the domain of $f$, then $M$ outputs longer and
longer prefixes of $f(x)$ when reading longer and longer prefixes of
$x$. This machine model has been defined in 
\cite[Chap.~2]{klaus}. If additionally one requires the existence of a
computable function $m \colon \mathbb{N}\rightarrow \mathbb{N}$ such that for
all $i\in\mathbb{N}$, for all infinite input $x$, $M$ writes at least $i$ output
symbols when reading $m(i)$ input symbols of $x$, we obtain the notion
of uniform computability. This offers promptness guarantees on
the production of output symbols with respect to the number of symbols
read on input.

Obviously, not all functions are computable. In this paper, we aim to solve the
following synthesis problem: given a (finite) specification of a
(partial) function $f$ from infinite words to infinite words, is $f$ computable
(respectively uniformly computable)? If it is the case, then the procedure
should return a Turing machine computing $f$  (respectively uniformly
computing $f$).

\subsection{Examples}

The function $S_{\sf swap}$ is computable. Since it is defined over all
inputs containing at least one $@$ symbol, if a Turing machine is fed
with such a word, it suffices for it to read its input 
and write it in working tape,
until the first $@$ symbol is met.
Now go back in the working tape to see the last symbol read before $@$, write the same in output tape.
It is followed by further going back to the beginning in the working tape, and copying all the symbols of working tape except the last one from left to right into the output tape.
This gives us the prefix $\sigma u_1$.
The suffix $@ u_2$ can simply be produced by copying the content of input tape into output tape.

Over the alphabet $\Sigma =
\{a,b\}$, consider the function $f_\infty$ defined by $f_\infty(u) =
a^\omega$ if $u$ contains infinitely many $a$s, and by $b^\omega$
otherwise. This simple function is not computable, as it requires to
read the whole infinite input to produce even the very first output
symbol.  For any word $u\in\Sigma^*$, we denote by
$\overline{u}$ its mirror (\eg $\overline{abaa} = aaba$). Consider the (partial) function $f_{\sf mir}$ 
defined on $(\Sigma^*\sharp)^\omega$ by $f(u_1\sharp u_2\sharp\ldots) =
\overline{u_1}\sharp \overline{u_2}\sharp\ldots$. It is computable by a machine
that stores its input $u_1$ in memory until the
first $\sharp$ is read, then outputs $\overline{u_1}$, and proceeds with $u_2$,
and so on. It is however not uniformly computable. Indeed, if it
were, with some $m \colon \mathbb{N}\rightarrow \mathbb{N}$, then, for inputs
$u_1\sharp u_2\sharp \cdots$ such that $|u_1|>m(1)$, it is impossible to
determine the first output symbol (which is the last of $u_1$) by
reading only a prefix of length $m(1)$ of $u_1$. 

Finally, consider the (partial) function $f_{\sf dbl}$ defined on
$(\Sigma^*\sharp )^\omega$ by $f(u_1\sharp u_2\sharp$ $\ldots) =$
$u_1u_1\sharp$ $u_2u_2$ $\sharp \ldots$. Similarly as before, it is computable but also
uniformly computable: to determine the $i^\text{th}$ output
symbol, it suffices to read an input prefix of length at most $i$. 
Indeed, let $u_1\sharp u_2\sharp \ldots \sharp u$ be a prefix of length $i$ of the
input  $x$, then $u_1u_1\sharp u_2u_2\sharp \ldots\sharp u$ is a prefix of $f(x)$ of
length  $\geq i$.

\subsection{Computability and continuity}

There are strong connections between computability
and continuity: computable functions are continuous
for the Cantor topology, that is where words are close to each other if
they share a long common prefix. Intuitively, it is because the very definition of
continuity asks that input words sharing longer and longer prefixes also share
longer and longer output prefixes. It is the case of the functions
$f_{\sf mir}$ and $f_{\sf dbl}$ seen before. Likewise, uniformly computable
functions are uniformly continuous. The reverse direction does not
hold in general:  
assuming an effective enumeration $M_1, M_2, \dots$ of Turing
machines (on finite word inputs),
the  function $f_{\sf halt}$ defined as $f_{\sf halt}:x\mapsto b_1b_2b_3 \ldots$ 
where $b_i \in \{0,1\}$ is such that $b_i=1$ if and only if $M_i$ halts
on input $\epsilon$, is not computable but (uniformly)
continuous (as it is constant).

\subsection{Beyond synchronous functions: regular functions}
Functional specifications in the Church $\omega$-regular synthesis problem range
over the class of \emph{synchronous} functions as described before:
they can be specified using automata over $\Sigma_i.\Sigma_o$. For
example, while $S_{\sf swap}$ and $f_\infty$ are synchronous, $f_{\sf mir}$ and $f_{\sf dbl}$ are not. 
In this paper, we intend to go much beyond this class by dropping the
synchronicity assumption and consider the so-called class of regular
functions. It is a well-behaved class, captured by several
models such as streaming $\omega$-string transducers (SST), deterministic two-way
Muller transducers with look around (\twoDFTla), and also 
by  MSO-transducers~\cite[Thm.~1, Prop.~1]{lics12}. We propose the model of deterministic two-way transducers 
with a prophetic \buchi{} look-ahead (\twoDFTpla) and show that they are equivalent to 
\twoDFTla. This kind of transducer is defined by a 
\emph{deterministic} two-way automaton without accepting states, extended with
output words on the transitions, and which can consult another
automaton, called the look-ahead automaton, to check whether an
infinite suffix satisfies some regular property. We assume this
automaton to be a prophetic B\"uchi automaton~\cite[Sec.~7]{DBLP:conf/birthday/CartonPP08}, because this class is naturally suited to implement a regular look-ahead, while capturing all regular
languages of infinite words. Look-ahead is necessary to capture
functions such as $f_\infty$. Two-wayness is needed to
capture, for instance, functions $f_{\sf dbl}$ and $f_{\sf mir}$.

\subsection{Contributions}

We call \emph{effectively reg-preserving functions} those
functions that effectively preserve regular languages by inverse
image~\cite{DBLP:journals/iandc/StearnsH63, 10.1145/1008304.1008306, DBLP:journals/tcs/SeiferasM76, PinS17}.
This includes for instance rational
functions, regular functions and the more general class of polyregular
functions~\cite{DBLP:journals/corr/abs-1810-08760, DBLP:journals/tcs/EngelfrietHS21}.
We first show that for
effectively reg-preserving functions, computability and
continuity coincide, respectively, uniform computability and uniform
continuity (Section~\ref{sec:contcomp}, Theorem~\ref{thm:cc}). To the best of our knowledge,
this connection was not made before. The connection is effective, in
the sense that when $f$ is effectively reg-preserving and
continuous (resp. uniformly continuous), we can effectively construct
a Turing machine computing $f$ (resp. uniformly computing $f$).

For rational functions (functions defined by
non-deterministic one-way B\"uchi transducers), we show that
continuity and uniform continuity are decidable in
\textsc{NLogSpace} (Section~\ref{sec:dec}, Theorem~\ref{thm:decrat}). Continuity and uniform continuity for rational functions were already known to be
decidable in \textsc{PTime}, from
Prieur~\cite[Prop.~4]{DBLP:journals/tcs/Prieur02}. However, Prieur's proof techniques do not
transfer to the two-way case.
We then prove that uniform continuity
(and hence uniform computability) is decidable in \textsc{PSpace} for regular functions given by
deterministic B\"uchi two-way transducers with look-ahead
(Section~\ref{sec:dec}, Theorem~\ref{thm:main1}). 
Using the same
techniques, we also get the decidability of
continuity (and hence computability) for regular functions
(Theorem~\ref{thm:main2}) in \textsc{PSpace}. For both problems, we
show that this upper-bound is tight, namely, both problems are
\textsc{PSpace}-hard.

Our proof technique relies on a characterisation
of non-continuous reg-preserving functions by the existence of pairs of sequences of words
which have a nice regular structure (Section~\ref{sec:contrat}, Lemma~\ref{lem:pairs}). Based on this, we derive
a decision procedure for continuity of rational functions by checking in
\textsc{NLogSpace} a structural transducer pattern. This pattern
expresses properties of synchronised loops on two runs of the
transducers. The case of
regular functions is much more involved as loops in two-way automata
or transducers are more complex. We also give a characterisation via a
structural pattern which we show to be checkable by a bounded-visit
non-deterministic two-way Parikh automaton of polynomial size,
yielding the \textsc{PSpace} upper bound as those automata can be
checked in \textsc{PSpace} for emptiness~\cite{DBLP:conf/fsttcs/FiliotGM19}.

\subsection{Related work}

The notion of
continuity has not been extensively studied in the transducers literature over infinite words. 
The work by Prieur~\cite{DBLP:journals/tcs/Prieur02} is the closest to ours, while 
\cite{DBLP:conf/lics/ChaudhuriSV13} looks at continuity of regular functions encoded by $\omega$-automata.

Notions of continuity with respect to language varieties have been studied for rational functions
of \emph{finite words} in~\cite{DBLP:conf/icalp/CadilhacCP17, DBLP:journals/lmcs/CadilhacCP19}. Our notion of uniform continuity can be linked to continuity with respect to a particular language variety which was \emph{not} studied in~\cite{DBLP:conf/icalp/CadilhacCP17} (namely the non-erasing variety generated by languages of the shape $uA^*$).
A quite strong Lipschitz continuity notion, called \emph{bounded variation} due to Choffrut (\eg\cite{DBLP:journals/tcs/Choffrut03}), was shown to capture, over finite words, the sequential functions (the corresponding topology is however trivial, hence simple continuity is not very interesting in this context).

Another result connecting computability and continuity is
from~\cite{DBLP:conf/mfcs/CadilhacKLP15} where the authors find that some notion of
computability by $\mathrm{AC}^0$ circuits corresponds, over sequential
functions, to continuity with respect to some language variety.

Our result on rational functions has been extended recently to
rational functions of infinite words over an infinite alphabet
in~\cite{DBLP:conf/fossacs/ExibardFR20}. More precisely, continuity for functions of infinite data
words defined by (one-way) transducers with registers has been shown
to be decidable. The proof of~\cite{DBLP:conf/fossacs/ExibardFR20} goes by reduction to the finite alphabet
setting and uses our result presented in the paper~\cite{DBLP:conf/concur/DaveFKL20}.

\paragraph{Changes from the conference version} 
The results in this paper are an extension of the work presented at CONCUR 2020~\cite{DBLP:conf/concur/DaveFKL20}. In the conference version,  
 Lemmas~\ref{lem:pairs} and~\ref{lem:pattern-rat} had only a short proof sketch;  this version contains the full technical details. A major change from the conference version is in 
Section~\ref{subsec:reg}. This has an added contribution, where we present  a new proof  for the decidability of continuity and uniform continuity for regular functions.
To aid us in this, we define what is called a  \emph{critical pattern} (Section~\ref{subsec:crit-pat}) and show that the existence of such patterns in  two way transducers help in deciding the continuity of the corresponding regular function (Proposition~\ref{prop:cont-crit-pat}).
Finally,  in the proof of Theorem~\ref{thm:main1} and Theorem~\ref{thm:main2}, we
analyze the complexity  of the pattern detection, which in turn yields
a (tight) upper-bound for the complexity of checking continuity and
uniform continuity of regular functions; this was left open in
the CONCUR 20 paper.


\section{Languages, Automata and Transducers over $\omega$-Words}\label{sec:prelim}

Given a finite set $\Sigma$, we denote by $\Sigma^*$
(resp. $\Sigma^\omega$) the set of finite (resp. infinite) words over
$\Sigma$, and by $\Sigma^\infty$ the set of finite and infinite
words. 
Let $\Sigma^j$ represent the set of all words over $\Sigma$ with length $j$.
We denote by $|u|\in\mathbb{N}\cup \{\infty\}$ the length of
$u\in\Sigma^\infty$ (in particular $|u|=\infty$ if
$u\in\Sigma^\omega$). For a word $w=a_1a_2a_3\dots$, $w[{\colon}j]$ denotes the prefix
$a_1a_2 \dots a_j$ of $w$.  Let $w[j]$ denote $a_j$, the $j^{\text{th}}$ symbol
of $w$ and $w[j{\colon}]$ denote the suffix $a_{j+1} a_{j+2} \dots$ of $w$.
For a word $w$ and $i\leq j$, $w[i{\colon}j]$ denotes the factor of $w$ with
positions from $i$ to $j$, both included.
For two words $u,v\in\Sigma^\infty$,  $u \prefix v$ (resp. $u\prec v$)
denotes that $u$
is a prefix (resp. strict prefix) of $v$ (in particular if $u,v\in\Sigma^\omega$, $u\preceq
v$ if and only if $u=v$). %
For $u \in \Sigma^*$, let ${\uparrow}u$ denote the set of words $w \in
\Sigma^\infty$ having $u$ as prefix \ie$u \prefix w$. 
Let $\textsf{mismatch}$ be a function which takes two words, and returns a boolean value, denoted by $\textsf{mismatch}(u,v)$ for $u$ and
$v$; it returns true if there exists a position $i\leq |u|,|v|$ such that $u[i]\neq v[i]$, and returns false otherwise.  
The longest common prefix between two words $u$ and $v$ is denoted by $u \wedge v$ and their distance is defined as $d(u, v) = 0$ if $u = v$, and $2^{-|u \wedge v|}$ if $u \neq v$.

A B\"uchi automaton is a tuple $B=(Q, \Sigma,\delta, Q_0, F)$ consisting 
of a finite set of states $Q$, a finite alphabet $\Sigma$, a set $Q_0 \subseteq Q$ of initial states, 
a set $F \subseteq Q$ of accepting states, and 
a transition relation $\delta \subseteq Q \times \Sigma \times Q$. 
A run $\rho$ on a word $w=a_1a_2 \ldots \in \Sigma^{\omega}$ 
starting in a state $q_1$ in $B$ is an infinite sequence $q_1  \stackrel{a_1}{\rightarrow}  q_2 \stackrel{a_2}{\rightarrow} \dots 
$ such that $(q_i, a_i, q_{i+1}) \in \delta$ for all $i \in \Nn$.  
Let $\Inf(\rho)$ denote the set of states visited infinitely often along $\rho$. The run $\rho$ is a final run 
if $\Inf(\rho) \cap F \neq \emptyset$. A run is \emph{accepting} if
it is final and starts from an initial state. A word
$w\in\Sigma^\omega$ is accepted ($w \in L(B)$) if it has an
accepting run. A language $L$ of $\omega$-words is called $\omega$-\emph{regular}
if $L = L(B)$ for some B\"uchi automaton $B$.

An automaton is co-deterministic if any two final runs on any word $w$ are the same \cite[Sec.~7.1]{DBLP:conf/birthday/CartonPP08}.  
Likewise, an automaton is co-complete if every word has at least one final run.   A prophetic automaton $P=(Q_P, \Sigma, \delta_P,Q_0, F_P)$ is a B\"uchi automaton which is co-deterministic and co-complete. 
Equivalently, a \buchi{} automaton is  prophetic if and only if each word admits a unique final run. 
The states of the prophetic automaton partition $\Sigma^{\omega}$: each state $q$ defines a set 
of words $w$ such that $w$ has a final run starting from $q$. For any state $q$, let $L(P,q)$ be the set  
of words having a final run starting at $q$. Then $\Sigma^{\omega}=\uplus_{q \in Q_P} L(P,q)$. 
It is known \cite[Thm.~7.2]{DBLP:conf/birthday/CartonPP08} that prophetic Büchi automata capture $\omega$-regular languages.

\subsection{Transducers}

We recall the definitions of one-way and two-way transducers over infinite words. 
A one-way transducer $\Aa$ is a tuple $(Q, \Sigma, \Gamma, \delta, Q_0, F)$ where $Q$ is a finite set of states, $Q_0, F$ respectively are sets of initial and  final states; $\Sigma, \Gamma$ respectively are the input and output alphabets;  $\delta \subseteq (Q \times \Sigma \times Q \times \Gamma^*)$ is the transition relation. We equip $\Aa$ with a \buchi{} acceptance condition. 
A transition in $\delta$ of the form $(q, a, q', \gamma)$ represents that from state $q$, on reading a symbol $a$, 
the transducer moves to state $q'$, producing the output $\gamma$. Runs, final runs and accepting runs are defined 
exactly as in B\"uchi automata, with the addition that each transition produces 
some output $\in \Gamma^*$.   

The output produced by a run $\rho$, denoted $\out{\rho}$,  is obtained by concatenating the outputs generated by transitions along $\rho$.  
Let $\dom{\Aa}$ represent the language accepted by the 
underlying automaton of $\Aa$, ignoring the outputs.
The relation computed by $\Aa$ is defined as $\sem{\Aa} = \{(u, v) \in \Sigma^\omega \times \Gamma^{\omega} \mid u \in \dom{\Aa}, 
\rho~\text{is an accepting run of}~u,
\out{\rho} = v \}$.\footnote{We assume that final runs always produce infinite words, which can be enforced syntactically by a B\"uchi condition such that any input word produces non-empty output in a single loop execution containing B\"uchi accepting state.}
We say that $\Aa$ is functional if $\sem{\Aa}$ is a function.
A relation (function) is \emph{rational} if it is recognised by a
one-way (functional) transducer.

Two-way transducers extend one-way transducers and two-way finite state automata. A two-way transducer 
is a two-way automaton with outputs. 
In \cite[Prop.~1]{lics12},  
regular functions are shown to be those definable by a  two-way deterministic transducer with Muller acceptance condition, 
along with a  regular look-around 
(\twoDFTla).
In this paper,  we propose  an alternative machine model for regular functions, namely, \twoDFTpla.  
A \twoDFTpla is a deterministic two-way automaton with outputs,  along with a look-ahead given by a
prophetic automaton. 

Let $\Sigma_{\leftend}=\Sigma \uplus \{\leftend\}$. 
Formally,  a \twoDFTpla is a pair $(\Tt, A)$ where $A=(Q_A, \Sigma, \delta_A, S_A, F_A)$ is a prophetic 
\buchi{}  automaton 
and  
$\Tt= (Q, \Sigma, \Gamma, \delta,$ $ q_0)$ is a two-way transducer  
such that $\Sigma$ and $\Gamma$ are  
finite input and output alphabets, $Q$ is a finite set of states, $q_0 \in Q$ is a unique initial state, 
$\delta\colon Q \times \Sigma_{\leftend} \times Q_A \rightarrow Q \times \Gamma^* \times \{-1,+1\}$ is a partial transition function. $\Tt$ has no acceptance condition: every infinite run in $\Tt$ is a final run.
A two-way transducer stores its input ${\leftend} a_1 a_2\dots$  on a two-way tape, and each index of the input can be read multiple times. 
A configuration of a two-way transducer is a tuple $(q, i) \in Q \times \mathbb{N}$ where $q \in Q$ is a state and 
$i \in \mathbb{N}$ is  the current position on the input tape. The position is an integer representing the gap between consecutive symbols. 
Thus, before ${\leftend}$, the position is 0, between ${\leftend}$ and  $a_1$, the position is 1, 
between $a_i$ and  $a_{i+1}$, the position is $i+1$ and so on. 
The \twoDFTpla is deterministic: for every word $w=~{\vdash} a_1a_2a_3 \ldots \in ~{\vdash} \Sigma^{\omega}$, every input position $i \in \mathbb{N}$, 
and state $q \in Q$, there is a unique state $p \in Q_A$ 
such that $a_i a_{i+1} \ldots \in L(A,p)$. 
Given $w=a_1a_2 \ldots$, from 
a configuration  $(q,i)$,  on a transition $\delta(q,a_i,p)=(q',\gamma,d)$, $d \in \{-1,+1\}$,  such that 
$a_{i+1} a_{i+2} \ldots \in L(A,p)$,   
we obtain the configuration $(q', i+d)$  and the output $\gamma$ is appended to the output produced so far. 
This transition is denoted as $(q,i) \stackrel{a_i,p/\gamma}{\longrightarrow} (q', i+d)$.
A run $\rho$ of a \twoDFTpla $(\Tt,A)$ is a sequence of transitions 
$(q_0,i_0=0) \stackrel{a_{i_0},p_1/\gamma_1}{\longrightarrow} (q_1, i_1)\stackrel{a_{i_1},p_2/\gamma_2}{\longrightarrow} \cdots $. 
The output of $\rho$, denoted $\out{\rho}$ is then $\gamma_1 \gamma_2 \cdots$.   	
The run $\rho$ reads the whole word $w$ if 
$\mathsf{sup}\{i_n \mid 0\leq n <|\rho|\} = \infty$. 
The output $\sem{(\Tt,A)}(w)$ 
of a word $w$ on run $\rho$ is defined 
only when $\mathsf{sup}\{i_n \mid 0\leq n <|\rho|\} = \infty$,
and equals $\out{\rho}$. 
\twoDFTpla are equivalent to \twoDFTla, and 
capture all regular functions. 
It is easy to see because different representations of $\omega$- look-ahead are equiexpressive, the look-behind of $\twoDFTla$ can be eliminated while preserving the expressiveness, and one can encode the acceptance condition of $\twoDFTla$ as part of the look-ahead in $\twoDFTpla$ 
(see Appendix~\ref{app:equitrans} for the proof).  	
\begin{thm}\label{thm:equivtrans}
	A  function $f\colon\Sigma^\omega\rightarrow \Gamma^\omega$ is regular if and only if it is \twoDFTpla definable. 
\end{thm}

\begin{figure}
	\tikzstyle{trans}=[-latex, rounded corners]
	\begin{center}
		\scalebox{.9}{
			
			\begin{tikzpicture}[->,>=stealth',shorten >=1pt,auto, semithick,scale=.7]
				\tikzstyle{every state}=[fill=gold]
				\node[state] at (-1,4) (B) {$q_1$} ;
				\node[state] at (3,4) (C) {$q_2$} ;
				\node[state,initial,initial where=left,initial text={}] at (-5,4) (A) {$q_0$} ;
				\node[state] at (3,0) (D) {$q_3$} ;
				\node[state] at (-5,0) (E) {$q_4$} ;
				
				\path(A)  edge node[above] {$\vdash, {p_1} \mid \epsilon, +1 $}
				(B);
				
				\path(B) edge[loop above] node
				{$\alpha, p_1 \mid  \alpha, +1 $}  (B);
				
				\path(B)  edge node[above] {$a, {p_2} \mid \epsilon, -1 $}
				(C);
				
				\path(C) edge[loop above] node
				{$\alpha \mid  \epsilon, -1 $}  (C);
				
				\path(C)  edge node[left] {$\vdash \mid \epsilon, +1 $}
				(D);
				
				\path(D) edge[loop below] node
				{$\alpha, {p_1} \mid  \alpha, +1 $}  (D);
				
				\path(D)  edge node[above] {$a, {p_2} \mid \epsilon, +1 $}
				(E);
				
				\path(E) edge[loop below] node {$b \mid b, +1$} (E);

				\path(A) edge node[right, align=left] {$\vdash,{p_2} \mid \epsilon, +1$} (E);
				
				\node[state] at (5.5, 4) (P1) {$p_1$} ;
				\node[state, accepting] at (10.5,4) (P3) {$p_3$} ;
				\node[state] at (8, 4) (P2) {$p_2$} ;
								
				\node[state] at (5.5,0) (P4) {$p_4$} ;
				\node[state, accepting] at (10.5,0) (P5) {$p_5$} ;	
				
				\path(P1) edge[loop above] node {$a,b$} (P1);
				\path(P1) edge node[above] {$a, b$} (P2);
				\path(P2) edge node[above] {$a$} (P3);
				\path(P3) edge[loop above] node {$b$} (P3);
								
				\path(P4) edge[loop above] node {$b$} (P4);
				\path(P4) edge [bend left=25] node[above] {$b$} (P5);
				\path(P5) edge[loop above] node {$a$} (P5);
				\path(P5) edge [bend left=25] node[below] {$a$} (P4);
				
			\end{tikzpicture}
		}
	\end{center}
	\caption{A \twoDFTpla with automaton on the right implementing the look-ahead.}
	\label{fig:func-g}
\end{figure}

\begin{exa}
	Consider the function $g\colon \Sigma^\omega \rightarrow \Gamma^\omega$ over $\Sigma = \Gamma = \{a, b\}$ such that $g(uab^\omega) = uub^\omega$ for $u \in \Sigma^*$ and $g(b^\omega) = b^\omega$. 
	The \twoDFTpla  is shown in Figure~\ref{fig:func-g} with the prophetic look-ahead automaton $A$ on the right. 
	The transitions are decorated as $\alpha, p\mid \gamma, d$ where $\alpha \in \{a,b\}$, $p$ is a state of $A$, $\gamma$ is the output and $d$ is the direction. 
	In transitions not using the look-ahead information, 
	the decoration is simply $\alpha \mid \gamma, d$. 
	Notice that $\Lang(A, p_1) = ~\Sigma^{+}ab^{\omega}$, $\Lang(A, p_2) = ~{a}b^\omega$, $\Lang(A, p_3) = b^\omega$, $\Lang(A, \{p_4, p_5\})$ is the set of words having infinitely many $a$s. Each word in $\Sigma^\omega$ has a unique final run.

\end{exa}
We also use a look-ahead-free version of two-way transducers in some of the proofs, where we also have a \buchi{} acceptance condition given by a set of states $F$, just as for \buchi{} automata.
The resulting model is called two-way deterministic \buchi{} transducer (\twoDBT).
The definitions of configuration, run, and the semantics are done just like for \twoDFTpla.

\section{Computability versus Continuity}\label{sec:contcomp}
Computability of a function on infinite words can be described intuitively in the following way: there is an algorithm which, given access to the input word, can enumerate the letters in the output word. We also investigate a stronger notion of computability, which we call \emph{uniform computability}. The main idea is that given some input word $x$ and some position $j$ one can compute the $j^{\text{th}}$ position of the output in time that depends on $j$ but not on $x$.
An appealing aspect of uniform computability is that it offers a uniform bound on the number of input symbols one needs to read in order to produce the output at some fixed precision.

\begin{defi}[Computability/Uniform computability]\label{def:computability}
	A function $f\colon \Sigma^{\omega} \rightarrow \Gamma^{\omega}$ is \emph{computable} if there exists a deterministic multitape Turing machine $M$ computing it in the following sense. 
	The machine $M$ has  a read-only one-way input tape, 
	a two-way working tape, and a write-only one-way output tape. All tapes have a left delimiter $\leftend$ and are infinite to the right.
	Let $x \in \dom{f}$. For any $j \in \mathbb{N}$, let $M(x,j)$ denote the output produced by $M$ 
	till the time it moves to the right of position $j$, onto position
	$j+1$ in the input (or $\epsilon$ if this move never happens).  
	The function $f$ is computable by $M$ if  
	for all $x \in \dom{f}$,  for all $i \geq 0$, there exists $j \geq 0$ 
	such that $f(x)[{\colon}i] \prefix M(x,j)$.
	
	Moreover if there exists a computable function $m \colon
	\mathbb{N}\rightarrow \mathbb{N}$ (called a \emph{modulus of continuity} for $M$)
	such that for all $x \in \dom{f}$, for all $i \geq 0$,  $f(x)[{\colon}i] \prefix M(x,m(i))$, $f$ is called \emph{uniformly computable}.
\end{defi}

It turns out that there is a quite strong connection between computability and continuity of functions. In particular computable functions are always continuous. This can be seen intuitively since given a deterministic Turing machine, it must behave the same on the common prefixes of two words. Hence two words with a very long common prefix must have images by the machine that have a somewhat long common prefix. This connection also transfers to uniform computability and uniform continuity.
We start by formally defining continuity and uniform continuity.

\begin{defi}[Continuity/Uniform continuity]
	\label{def:continuity}
	\
	We interchangeably use the following two textbook definitions \cite{rudin} of continuity.
	\begin{enumerate}
		\item A function $f\colon\Sigma^\omega\rightarrow \Gamma^\omega$ is
		continuous at $x\in\dom{f}$ if (equivalently)
		\begin{itemize}
			\item[(a)]  for all $(x_n)_{n \in \Nn}$ converging to $x$,
			where $x_i \in \dom{f}$ for all $i \in \Nn$, $(f(x_n))_{n \in \Nn}$
			converges.
			\item[(b)] $\forall i \geq 0$  $\exists j \geq 0$ $\forall y \in \dom{f}$, $|x \wedge y| \geq j \Rightarrow |f(x) \wedge f(y)| \geq i$
		\end{itemize}
		\item A function is continuous if it is continuous at every $x\in\dom{f}$.
			
	\end{enumerate}
	
		\noindent A function $f\colon \Sigma^{\omega} \rightarrow \Gamma^{\omega}$ is uniformly continuous if:
		
		there exists $m\colon \mathbb N\rightarrow \mathbb N$, called a \emph{modulus of continuity} for $f$ such that,
		
		$\forall i \geq 0$, $\forall  x,y \in \dom{f}$, $|x \wedge y| \geq m(i) \Rightarrow |f(x) \wedge f(y)| \geq i$.  
\end{defi}

\begin{exa}
	As explained in the introduction, the function $f_\infty$ is not continuous, and, as we will see later, is thus not computable. 
	The function $f_{\textsf{halt}}$ is continuous, even uniformly continuous (it is constant) yet is obviously not computable.
	The function $f_{\textsf{mir}}$ is computable, however is not uniformly continuous, two words can be arbitrarily close but with far away outputs: consider $a^n\sharp^\omega$ and $a^nb\sharp^\omega$.
	Finally, the function $f_{\textsf{dbl}}$ is uniformly computable.
\end{exa}

We now investigate the relationship between continuity and computability for functions that are effectively reg-preserving.
More precisely, we
say that a function $f \colon \Sigma^\omega\rightarrow \Gamma^\omega$ is \emph{effectively reg-preserving}
if there is an algorithm which, for any automaton recognising a regular language $L\subseteq \Gamma^\omega$, produces an automaton recognising the language $f^{-1}(L) = \{ u\mid f(u)\in L\}$. 

Two well-studied classes (see \eg\cite{DBLP:conf/icla/Filiot15}) of reg-preserving functions are the rational and the regular functions, which we will study in Section~\ref{sec:dec}.
As announced, continuity and computability coincide for effectively reg-preserving functions:
\begin{thm}\label{thm:cc}
	An effectively reg-preserving function $f\colon \Sigma^\omega \to \Gamma^\omega$ is computable (resp. uniformly computable)  if and only if it is continuous (resp. uniformly continuous).
\end{thm}

\begin{proof}
	$\Rightarrow$) This implication is easy and actually holds without the reg-preserving assumption.
	Assume that $f$ is computable. 
	We prove the continuity of $f$. 
	Let $M$ be the machine computing $f$.   Let $x \in \dom{f}$ 
	be the content on the input tape of $M$. 
	Intuitively, the longer the prefix of input $x\in\dom{f}$ is processed by $M$,
	the longer the output produced by $M$ on that prefix, which converges
	to $f(x)$, according to the definition of computability. 
	For all $i \geq 0$,  define $\pref{M}{x}{i}$ to be the smallest $j \geq 0$ such that,  
	when $M$ moves to the right of $x[{\colon}j]$ into cell $j+1$, it has output 
	at least the first $i$ symbols of $f(x)$.   
	For any $i \geq 0$, choose $j = \pref{M}{x}{i}$. Consider any $z\in \dom{f}$ such that $|x \wedge z| \geq j$.
	After reading $j$ symbols of $x$, the machine $M$ outputs at least $i$ symbols of $f(x)$.
	Since $|x \wedge z| \geq j$, the first $j$ symbols of $x$ and $z$ are the same, and 
	$M$ being deterministic,  outputs the same $i$ symbols on reading the first $j$ symbols of $z$ as well. 
	These $i$ symbols form the prefix for both $f(x), f(z)$, and hence, 
	$|f(x) \wedge f(z)| \geq i$. 
	
	Thus, for every $x \in \dom{f}$ and for all $i$, there exists  
	$j= \pref{M}{x}{i}$ such that, for all $z \in \dom{f}$, 
	if  $|x \wedge z| \geq j$, then  
	$|f(x) \wedge f(z)| \geq i$ implying continuity of $f$. 
	Moreover, if $f$ is uniformly computable, then the modulus of continuity of $M$ is in particular a modulus of continuity for $f$ and is thus uniformly continuous.

	\begin{algorithm}
		\SetAlgoLined
		\SetKwFor{For}{for}{do}{}
		\KwIn{$x \in \Sigma^\omega$}
		\SetKwIF{If}{ElseIf}{Else}{if}{then}{else if}{else}{}
		\ou~{:=} $\epsilon$ ; ~~  \textcolor{red}{\tt{this is written on the working tape}} \\
		\For{$i = 0$ to  $+\infty$}{
			\For{$\gamma \in \Gamma$}{
					\If{$f( \upward{x[{\colon}i]})\mathrel{\subseteq} {\uparrow}$\ou$.\gamma$}{
					\ou~{:=} \ou.$\gamma$ ; \textcolor{red}{\tt{append to the working tape}} \\
					output $\gamma$ ; ~~  \textcolor{red}{\tt{this is written on the output tape}}
				}
			}
		}
		\caption{Algorithm describing $M$.}
		\label{algo:out}
	\end{algorithm}
	
	$\Leftarrow$) The converse direction is less trivial and makes use of the reg-preserving assumption. Suppose that $f$ is
	continuous. We design the machine $M$, represented as
	Algorithm~\ref{algo:out}, which is shown to compute $f$. This
	machine processes longer and longer prefixes $x[{\colon}i]$ of its
	input $x$ (for
	loop at line 2), and tests (line 4) whether a symbol $\gamma$ can
	be safely appended to the output.
	The test ensures that the invariant $\textsf{out}\preceq f(x)$ is preserved at any point.
	Moreover, the continuity of $f$ at $x$ ensures that $\textsf{out}$ is updated infinitely often.
	The only thing left  to obtain computability is that the test of line 4 is decidable.	
	Let $u$ and $v$ be two words,  deciding $f( {\uparrow}u)\subseteq {\uparrow} v$ is equivalent to deciding if $\dom{f}\cap {\uparrow} u \subseteq f^{-1}({\uparrow} v)$.
	These sets are effectively regular since $u,v$ are given and $f$ is effectively reg-preserving, and $\dom{f}=f^{-1}(\Gamma^\omega)$. Since the constructions are effective,
	and the languages are regular, the inclusion is decidable.
	
	We only have left to show that if $f$ is moreover uniformly continuous, then $M$ has a computable modulus of continuity.
	We start by showing that $f$ has a computable modulus of continuity.
	Let us consider the predicate $P(i,j)\colon \forall x,y\ |x\wedge y|\geq j \Rightarrow |f(x)\wedge f(y)|\geq i$. 
	Then we define $m\colon i\mapsto \min\set{j|\ P(i,j)}$. Since $f$ is uniformly continuous, $m$ is indeed well defined and is a modulus of continuity of $f$. To show that $m$ is computable, we only have to show that $P(i,j)$ is decidable.
	
	Let us consider the negation of $P(i,j)$: there exist $u,x_1,x_2,v_1,v_2,w_1,w_2$ such that $|u|=j$, and $f(ux_k)=v_kw_k$ for $k\in \set{1,2}$ with $|v_1|=|v_2|=i$ and $v_1\neq v_2$.
	Hence to decide $\neg P(i,j)$, we only have to find two words $v_1\neq v_2$ in $\Gamma^i$, such that 
	$S\neq\emptyset$ where 
	$S=\{vw \mid v \in \Sigma^j, ~\exists w_1, w_2, ~\text{s.t.}~ vw_1\in f^{-1}({\uparrow} v_1),~ vw_2 \in f^{-1}(\upward{v_2}) \}$.
	Since $f$ is effectively reg-preserving, $S$ is effectively regular. By searching exhaustively for words $v_1,v_2\in \Gamma^i$ we get decidability of $P(i,j)$.
	We only have left to define a modulus of continuity for $M$. Let $m' \colon \mathbb N\rightarrow \mathbb N$ be defined by $m'(i)=m(i)+i$. If we read $m(i)$ symbols, we know we can output at least $i$ symbols. 
	Hence in each of the next $i$ steps, we are guaranteed to output a letter. Hence $m'$ is a modulus of continuity for $M$ and $f$ is uniformly computable.
\end{proof}

\begin{rem}
	Note that we focus on functions that are effectively reg-preserving, but Algorithm~\ref{algo:out} is actually more general than that.
	The continuity-computability equivalence indeed carries over to any class of functions for which the test in line 4 is decidable.
\end{rem}

\section{A Characterisation of Continuity and Uniform Continuity}\label{sec:contrat}
We provide here a characterisation of continuity (and uniform continuity) for reg-preserving functions (we don't need effectiveness here).
The characterisation is based on a study of some particular properties of sequences and pairs of sequences which we define below.

\paragraph{Topology preliminaries} A \emph{regular word} (sometimes called ultimately periodic) over
$\Sigma$ is a word of the form $uv^\omega$ with $u\in \Sigma^*$ and
$v\in \Sigma^+$. The set of regular words is denoted by $\rat\Sigma$.
The \emph{topological closure} of a language $L\subseteq \Sigma^\omega$, denoted by $\bar L$, is
the smallest language containing it and closed under taking the limit
of converging sequences, \ie$\bar L = \set{x\mid\ \forall u\prec x,
	\exists y,\ uy\in L}$. A language is \emph{closed} if it is equal to its closure.
A sequence of words $\tuple{x_n}_{n\in\Nn}$ is called regular if
there exists $u,v,w\in \Sigma^*$, $z\in \Sigma^+$ such that for all
$n\in \Nn$, $x_n=uv^nwz^\omega$. 
The proof of the characterisation of continuity (Lemma~\ref{lem:pairs}) requires the following folklore or easy to show propositions:
\begin{prop}
	\label{prop:group}
	Let $L \subseteq \Sigma^\omega$ be regular language, then
	\begin{enumerate}
		\item\label{prop:rat-bar}  $\bar L$ is regular,
		\item\label{prop:rat-dense} $L\subseteq \overline{L\cap \rat \Sigma}$ (\ie the regular words are dense in
		a regular language),
		\item\label{prop:rat-seq}
		any regular word of $\bar L$ is the limit of a regular sequence of
		$L$
	\end{enumerate}
\end{prop}

\begin{proof}
	Proof of Proposition~\ref{prop:group}.\ref{prop:rat-seq}: Let $L\subseteq \Sigma^{\omega}$ be regular.
	Let $uv^\omega$ be a regular word in $\bar L$.
	By regularity of $L$, there is a power of $v$, $v^k$ such that for any words, $w,x$, $wv^kx\in L \Leftrightarrow wv^{2k}x\in L$.
	Let us consider the language $K=\set{x\mid\ uv^kx\in L}$ which is non-empty since $uv^\omega\in \bar L$. Moreover, $K$ is regular, and thus contains a regular word $wz^\omega$.
	Hence we have that $uv^\omega$ is the limit of the sequence $\tuple{uv^{kn}wz^\omega}_{n\in \Nn}$ of $L$.
\end{proof}

\begin{defi}
	Let $f\colon \Sigma^\omega\rightarrow\Gamma^\omega$.
	
	Let $\tuple{x_n}_{n\in \Nn}$ be a sequence of words in $\dom f$ converging to $x\in \Sigma^\omega$, such that $\tuple{f(x_n)}_{n\in \Nn}$ is not convergent. Such a sequence is called a \emph{bad sequence} at $x$ for $f$.

	Let $\tuple{x_n}_{n\in \Nn}$ and $\tuple{x_n'}_{n\in \Nn}$ be two sequences in $\dom f$ both converging to $x\in \Sigma^\omega$, such that either $\tuple{f(x_n)}_{n\in \Nn}$ is not convergent, $\tuple{f(x_n')}_{n\in \Nn}$ is not convergent, or $\lim_nf(x_n)\neq\lim_nf(x_n')$. Such a pair of sequences is called a \emph{bad pair} of sequences at $x$ for $f$.
	
	
	A pair of sequences is \emph{synchronised} if
	it is of the form:
	$\tuple{\tuple{uv^nwz^\omega}_n,\tuple{uv^nw'z'^\omega}_n}$
\end{defi}

\begin{prop}
	\label{prop:bad-pairs}
	A function is \emph{not} continuous if and only if it has a bad pair at some point of its domain.
	A function is \emph{not} uniformly continuous if and only if it has a bad pair.
\end{prop}
\begin{proof}
	The case of continuous functions is obtained just by definition. For
	uniform continuity, consider a function $f$ with a bad pair $(\tuple{x_n}_{n\in \Nn},\tuple{x_n'}_{n\in \Nn})$, and let us show that it is not uniformly continuous.
	We can assume that both $\tuple{f(x_n)}_{n\in \Nn}$ and $\tuple{f(x_n')}_{n\in \Nn}$ converge. Otherwise we can extract subsequences that converge, by compactness of $\Gamma^\omega$.
	Moreover, since the pair is bad, one can assume that they converge to different limits $y\neq y'$.
	Let $i$ be such that $y[i]\neq y'[i]$.
	For any $j$, one can find $N$ such that for all $n\geq N$, $|x_n \wedge x'_n|\geq j$ since both 
	sequences converge to $x$.
	Since $\tuple{f(x_n)}_{n\in \Nn}$ converges to $y$, we can ensure that $N$ is large enough so that for all $n\geq N$, $|f(x_n)\wedge y|\geq i$.
	We can also ensure that for $n \geq N$, $|f(x_n') \wedge y'| \geq i$ holds.
	Let $n\geq N$, we have both $|x_n \wedge x'_n|\geq j$ and $|f(x_n)\wedge f(x_n')|<i$, which means that $f$ is not uniformly continuous.
	
	Let $f$ be a function which is not uniformly continuous, we want to exhibit a bad pair of $f$.
	According to the definition, there exists $i$ such that for all $j$ there exist $x_j, x_j'$ with $|x_j\wedge x'_j|\geq j$ but $|f(x_j)\wedge f(x_j')| < i$.
	By compactness of $\Sigma^\omega$, there exists a subsequence of $\tuple{x_j}_{j\in \Nn}$ which is convergent.
	Let $\tuple{x_{\tau(j)}}_{j\in \Nn}$, with $\tau \colon \mathbb N \rightarrow \mathbb N$ increasing, denote such a subsequence.
	Then we have for all $j$ that $|x_{\tau(j)} \wedge x_{\tau(j)}'|\geq \tau(j)\geq j$ and $|f(x_{\tau(j)})\wedge f(x_{\tau(j)}')| < i$.
	Therefore up to renaming the sequences, we can assume that for all $j$, $|x_j\wedge x_j'|\geq j$ and $|f(x_j)\wedge f(x_j')| < i$, with $\tuple{x_j}_{j\in \Nn}$ being convergent.
	By repeating the process of extracting subsequences, we can assume that $\tuple{x_j'}_{j\in \Nn}$, $\tuple{f(x_j)}_{j\in \Nn}$, $\tuple{f(x_j')}_{j\in \Nn}$ are also convergent.
	Since for any $j$, $|x_j\wedge x_j'|\geq j$, the two sequences converge to the same limit. In the end we obtain that $(\tuple{x_j}_{j\in \Nn},\tuple{x_j'}_{j\in \Nn})$ is a bad pair for $f$ at $\lim_j x_j=\lim_jx_j'$.
	
	Note that in case $\dom{f}$ is not compact, then we may not have a subsequence of $\tuple{x_j}_{j\in \Nn}$ which converges in $\dom{f}$. However, the definition of bad pairs does not require convergence in the domain; it only asks for convergence to some $x$, which need not be in $\dom{f}$. 
\end{proof}

The main result of this section is the following lemma which says that one can restrict to considering only \emph{synchronised} bad pairs. In the following sections this characterisation will be used to decide continuity/uniform continuity.

\begin{lem}[Characterisation]
	\label{lem:pairs}
	A  reg-preserving function is not
	continuous  if and only if it has a synchronised
	bad pair at some point of its domain. 
	A  reg-preserving function is not
	uniformly continuous  if and only if it has a synchronised
	bad pair. 
\end{lem}

\begin{proof}
	This lemma extends Proposition~\ref{prop:bad-pairs}. It shows that, in the case of reg-preserving functions, one can restrict to considering synchronised pairs, which are much easier to deal with.
	The only if direction is a simple consequence of Proposition~\ref{prop:bad-pairs}. For the other direction,
	 we first show that for a reg-preserving function $f$, if there is a bad pair at some $x$, then there is one at some \emph{regular} $z$, \ie$z=uv^\omega$ for some finite words $u,v$.
	Moreover, for the case of non-uniform continuity, we show that $z$ can be chosen so that $x\in \dom f \Leftrightarrow z\in \dom f$.
	In a second step, since we have two sequences converging to a regular $z$, we show how to replace the bad pair by a bad pair of \emph{regular} sequences, still using the fact that $f$ is reg-preserving. Finally, we prove that we can \emph{synchronise} these two regular sequences and end up with a synchronised bad pair at $z$.
	
	Let $f\colon \Sigma^\omega\rightarrow\Gamma^\omega$ be a reg-preserving function which is not uniformly continuous.
	Then according to Proposition~\ref{prop:bad-pairs}, it has a bad pair $(\tuple{x_n}_{n\in \Nn},\tuple{x_n'}_{n\in \Nn})$ at some point $x$. If $f$ is not continuous then we can assume that $x\in \dom f$.
	Our goal is to exhibit a synchronised bad pair at a point $z$, such that if $x\in \dom f$ then $z \in \dom f$.
	
	By compactness of $\Gamma^\omega$, we can extract subsequences such that both $\tuple{f(x_n)}_{n\in \Nn}$ and $\tuple{f(x_n')}_{n\in \Nn}$ converge.
	Moreover, since the pair is bad, one can assume that they converge to different limits $y\neq y'$.
	Let $i$ be such that $y[i]\neq y[i']$ and let $B_y=\set{z\mid\ |y\wedge z|\geq i}={\uparrow}y[{\colon}i]$ and $B_{y'}=\set{z\mid\ |y'\wedge z|\geq i}={\uparrow}y'[{\colon}i]$.
	By definition we have $B_y\cap B_{y'}=\emptyset$, and moreover both sets $B_{y},B_{y'}$ are regular.
	Up to extracting subsequences, we can assume that for all $n$, $x_n\in f^{-1}(B_y)$ and $x_n'\in f^{-1}(B_{y'})$. This means that $x\in \overline{f^{-1}(B_{y})}\cap \overline{f^{-1}(B_{y'})}$.
	Since $f$ is reg-preserving, and from
	Proposition~\ref{prop:group}.\ref{prop:rat-bar}, the set
	$\overline{f^{-1}(B_{y})}\cap \overline{f^{-1}(B_{y'})}$ is regular,
	and non-empty. Hence there exists a regular word $z\in
	\overline{f^{-1}(B_{y})}\cap \overline{f^{-1}(B_{y'})}$. 
	Moreover, since $\dom f$ is also regular, we can choose $z$ so that $x\in \dom f  \Leftrightarrow z\in\dom f $.
	Since $z\in \overline{f^{-1}(B_{y})}$, there is a sequence
	$\tuple{z_n}_{n\in \Nn}$ of words in $f^{-1}(B_{y})$ which converges
	to $z$. Furthermore, since $f^{-1}(B_{y})$ is regular and since $z$ is
	a regular word, we can assume, from
	Proposition~\ref{prop:group}.\ref{prop:rat-seq}, that the sequence
	$\tuple{z_n}_{n\in \Nn}$ is regular. Similarly, there is a regular
	sequence $\tuple{z_n'}_{n\in \Nn}$ in $f^{-1}(B_{y'})$ which converges
	to $z$.

	If either sequence $\tuple{f(z_n)}_{n\in \Nn}$ or
	$\tuple{f(z_n')}_{n\in \Nn}$ is not convergent then we are done, because one of the pairs
	$(\tuple{z_n}_{n\in \Nn},\tuple{z_n}_{n\in \Nn})$ or $(\tuple{z_n'}_{n\in \Nn},\tuple{z_n'}_{n\in \Nn})$ is bad and of course synchronised.
	If both sequences are convergent, then $\lim_n f(z_n) \in B_y$ and
	$\lim_n f(z_n') \in B_{y'}$ (because $B_y$ and $B_{y'}$ are both
	closed), which means that $|\lim_n f(z_n)\wedge \lim_n f(z_n')|\leq
	i$, hence the pair $\tuple{\tuple{f(z_n)}_{n\in
			\Nn},\tuple{f(z_n')}_{n\in \Nn}}$ is bad.
	
	We only have left to show that we can synchronise the sequences.
	Let $z_n=uv^nwt^\omega$ and let $z'_n=u'v'^nw't'^\omega$
	
	Let us first assume that $z'_n$ is constant equal to $z$. Let $z'=u^{-1}z$, we have that $v^nz'=z'$ and $uv^nz'=z$ for any $n$.
	Thus the pair $\tuple{\tuple{z_n}_{n\in \Nn},\tuple{z}_{n\in \Nn}}$ is a synchronised bad pair.
	
	Let us now assume that neither sequence is constant, which means that $|v|,|v'|>0$. Without loss of generality, let us assume that $|u|\geq |u'|$, let $k\in \Nn$, let $p<|v'|$ be such that $|u'|+k|v'|+p=|u|$, let $v''=v'[p+1\colon|v'|]v'[1\colon p]$ and let $w''=v'[p+1\colon |v'|]w'$.
	Then we can write $z''_n=z'_{n+k+1}=u'(v')^kv'[1\colon p]\cdot(v'')^n \cdot w''t'^\omega=uv''^nw''t'^\omega$.
	
	Note that $v^\omega=v''^\omega$, which means that $v^{|v''|}=v''^{|v|}$.
	Let $y_n=z_{|v''|n}{=}uv^{|v''|n}wt^{\omega}$ and let $y''_n=z''_{|v|n}{=}u{v''}^{|v|n}w't'^{\omega}$.
	Then the pair $\tuple{\tuple{y_n}_{n\in \Nn},\tuple{y''_n}_{n\in \Nn}}$ is a synchronised bad pair at $z$.
	
\end{proof}

\section{Deciding Continuity and Uniform Continuity}\label{sec:dec}
We first show how to decide (uniform) continuity for rational and then
for regular functions.

\subsection{Rational case}

We exhibit structural patterns which are
shown to be satisfied by a one-way B\"uchi transducer if and only if the rational
function it defines is not continuous (resp. not uniformly
continuous). We express those patterns in the \emph{pattern logic} defined in \cite[Sec.~6]{DBLP:conf/dlt/FiliotMR18}, which is based on
existential run quantifiers of the form $\exists \pi \colon
p\xrightarrow{u|v} q$ where $\pi$ is a run variable, $p,q$ are state variables
and $u,v$ are word variables. Intuitively,  there
exists a run $\pi$ from state $p$ to state $q$ on input $u$, producing
output $v$. 
A one-way transducer is called \emph{trim} if each of its states appears in
some accepting run. Any one-way B\"uchi transducer can be trimmed in polynomial
time.
The structural patterns for trim transducers are given in Figures~\ref{fig:pattern-1} and~\ref{fig:pattern-2}. 
The predicate
$\textsf{init}(p)$ expresses that $p$ is initial while
$\textsf{acc}(p)$ expresses that it is accepting. The predicate
$\textsf{mismatch}$ expresses the existence of a mismatch between
two words, as
defined in Section~\ref{sec:prelim}.
\begin{figure}[t]
	\begin{minipage}{0.65\linewidth}
		\begin{footnotesize}
			$\phi_{\text{cont}}=\begin{array}{l}
			\exists \pi_1\colon p_1 \xrightarrow{u|u_1} q_1,\ \exists \pi_1'\colon q_1 \xrightarrow{v|v_1} q_1\\
			\exists \pi_2\colon p_2 \xrightarrow{u|u_2} q_2,\ \exists \pi_2'\colon q_2 \xrightarrow{v|v_2} q_2,\ \exists \pi_2''\colon q_2 \xrightarrow{w|w_2} r_2\\
			\big(\mathsf{init}(p_1)\wedge \mathsf{init}(p_2)\wedge\mathsf{acc}(q_1)\big)\wedge\\
			\big( \mismatch(u_1,u_2)\vee(v_2=\epsilon\wedge \mismatch(u_1,u_2w_2)) \big)
			\end{array}$
		\end{footnotesize}
	\end{minipage} 
	\begin{minipage}{0.3\linewidth}
\tikzstyle{trans}=[-latex, rounded corners]
\begin{center}
	\scalebox{0.6}{
		\begin{tikzpicture}[->,>=stealth',shorten >=1pt,auto, semithick,scale=.8]
		\tikzstyle{every state}=[fill=gold]
		\node[state,initial,initial where=left,initial text={}] at (-4,4) (A) {$p_1$} ;
		\node[state, accepting] at (-1,4) (B) {$q_1$} ;
		
		\node[state,initial,initial where=left,initial text={}] at (-4,1.5) (C) {$p_2$} ;
		\node[state] at (-1,1.5) (D) {$q_2$} ;
		\node[state] at (2,1.5) (E) {$r_2$} ;
		
		\path(A) edge node[above] {$u | u_1 $}
		(B);
		
		\path(B) edge[loop above] node {$v | v_1 $}  (B);
		
		\path(C) edge node[above] {$u | u_2 $} (D);
		
		\path(D) edge [loop above] node {$v | v_2 $} (D);
		
		\path(D) edge node[above] {$w | w_2 $} (E);
		
		\end{tikzpicture}
	}
\end{center}
	\end{minipage}
	\caption{Pattern characterising non-continuity of rational
		functions given by \emph{trim} one-way B\"uchi transducers. 
		\label{fig:pattern-1}}
\end{figure}

\begin{figure}[t]
	\begin{minipage}{0.63\linewidth}
		\begin{footnotesize}
			$\phi_{\text{u-cont}}=\begin{array}{lll}
			\exists \pi_1\colon p_1 \xrightarrow{u|u_1} q_1,\ \exists \pi_1'\colon q_1 \xrightarrow{v|v_1} q_1,\ \exists \pi_1''\colon q_1 \xrightarrow{w|w_1} r_1\\
			\exists \pi_2\colon p_2 \xrightarrow{u|u_2} q_2,\ \exists \pi_2'\colon q_2 \xrightarrow{v|v_2} q_2,\ \exists \pi_2''\colon q_2 \xrightarrow{w|w_2} r_2\\
			\big(\mathsf{init}(p_1)\wedge \mathsf{init}(p_2)\big)\wedge\\
			\big( \mismatch(u_1,u_2)\vee(v_1=\epsilon\wedge \mismatch(u_1w_1,u_2))\\
			\hspace{1.3cm}  \vee(v_1=v_2=\epsilon\wedge \mismatch(u_1w_1,u_2w_2)) \big)
			\end{array}
			$
		\end{footnotesize}
	\end{minipage} 
	\begin{minipage}{0.35\linewidth}

\tikzstyle{trans}=[-latex, rounded corners]
\begin{center}
	\scalebox{0.6}{
		
		\begin{tikzpicture}[->,>=stealth',shorten >=1pt,auto, semithick,scale=.8]
		\tikzstyle{every state}=[fill=gold]
		\node[state,initial,initial where=left,initial text={}] at (-4,4) (A) {$p_1$} ;
		\node[state] at (-1,4) (B) {$q_1$} ;
		\node[state] at (2,4) (C) {$r_1$} ;
		\node[state,initial,initial where=left,initial text={}] at (-4,1.5) (D) {$p_2$} ;
		\node[state] at (-1,1.5) (E) {$q_2$} ;
		\node[state] at (2,1.5) (F) {$r_2$} ;
		
		\path(A) edge node[above] {$u | u_1 $} (B);
		
		\path(B) edge[loop above] node {$v | v_1 $}  (B);
		
		\path(B) edge node[above] {$w | w_1 $} (C);
		
		\path(D) edge node[above] {$u | u_2 $} (E);
		
		\path(E) edge [loop above] node {$v | v_2 $} (E);
		
		\path(E) edge node[above] {$w | w_2 $} (F);
		
		\end{tikzpicture}
	}
\end{center}
\end{minipage}
	
	\caption{Pattern characterising non-\emph{uniform} continuity of rational
		functions given by \emph{trim} one-way B\"uchi transducers.\label{fig:pattern-2}} 
\end{figure}

\begin{lem}
	\label{lem:pattern-rat}
	A trim one-way B\"uchi transducer defines a non-continuous (resp. non-uniformly continuous) function if and only if it satisfies the formula
	$\phi_{\text{cont}}$ of Figure~\ref{fig:pattern-1} (resp. the formula
	$\phi_{\text{u-cont}}$ of Figure~\ref{fig:pattern-2}).
\end{lem}
\begin{proof}
	
	Showing that the patterns of Figure~\ref{fig:pattern-1} and Figure~\ref{fig:pattern-2} induce
	non-continuity and non-uniform continuity, respectively, is quite
	simple. 
	Indeed, the first pattern $\phi_{\text{cont}}$ is a witness that
	$\tuple{uv^nwz}_{n\in \Nn}$ is a bad sequence at a point 
	$uv^\omega$ of its domain, for $z$ a word with a final run from $r_2$,
	which entails non-continuity by Proposition~\ref{prop:bad-pairs} (if a sequence $s$ is bad then $(s,s)$ is bad). Similarly, the pattern $\phi_{\text{u-cont}}$ witnesses that the
	pair $\tuple {\tuple{uv^nwz}_{n\in \Nn},\tuple{uv^nw'z'}_{n\in \Nn}}$
	is synchronised and bad (with $z,z'$ words that have a final run from
	$r_1,r_2$, respectively), which entails non-uniform continuity.
	
	For the other direction, we again use Lemma~\ref{lem:pairs}.
	From a synchronised bad pair, we can find a pair of runs with a
	synchronised loop, such that iterating the loop does not affect the
	existing mismatch between the outputs of the two runs, which is in
	essence what the pattern formulas of Figure~\ref{fig:pattern-1} and
	Figure~\ref{fig:pattern-2} state. 
	Let us consider the continuous case formally. The uniformly continuous case is similar.
	Let $f$ be a function realised by a trim transducer $T$.
	
	Let us first show that if the pattern of $\phi_{\text{cont}}$ appears, then we can exhibit a bad pair at a point of the domain. 
	Let us assume $\phi_{\text{cont}}$ holds.
	Then the word $uv^\omega$ is in the domain of the function realised by the transducer since it has an accepting run.
	Let $zt^\omega$ be a word accepted from state $r_2$, which exists by trimness, and let us consider the pair $\tuple{\tuple{uv^\omega}_{n\in \Nn},\tuple{uv^nwzt^\omega}_{n\in \Nn}}$.
	We have $f(uv^\omega)=u_1v_1^\omega$ and $u_2\prefix \lim_n f(uv^nwzt^\omega)$. If $\mismatch(u_1,u_2)$ then the pair is bad, otherwise, we must have $v_2=\epsilon$ and thus $u_2w_2\prefix \lim_n f(uv^nwzt^\omega)$. Since $\mismatch(u_1,u_2w_2)$ we again have that the pair is bad.
	
	Let us assume that the function $f$ is not continuous at some point $x\in\dom{f}$.
	Then according to Lemma~\ref{lem:pairs}, there is a synchronised bad pair $\tuple{\tuple{uv^nwz^\omega}_{n\in \Nn},\tuple{uv^nw'z'^\omega}_{n\in \Nn}}$ converging to some point of $\dom f$.
	
	Our goal is to exhibit a pattern as in $\phi_{\text{cont}}$.
	If $v$ is empty, then the sequences are constant which contradicts the functionality of $T$.
	So both the sequences converge to $uv^\omega$, but their output sequences either do not converge or if converges, then not to $f(uv^\omega)$.
	Because of the mismatch, either $\tuple{\tuple{uv^nwz^\omega}_{n\in \Nn}, \tuple{uv^\omega}_{n\in \Nn}}$ is a bad pair or 
	$\tuple{\tuple{uv^\omega}_{n\in \Nn},\tuple{uv^nw'z'^\omega}_{n\in \Nn}}$ is a bad pair (or both).
	Without loss of generality, let $\tuple{\tuple{uv^nwz^\omega}_{n\in \Nn},\tuple{uv^\omega}_{n\in \Nn}}$ be  a bad pair.
	
	Let us consider an accepting run $\rho$ of $T$ over $uv^\omega$. Since
	the run is accepting, there is an accepting state $q_1$ visited infinitely often. Let $l\in\set{1,\ldots,|v|}$ be such that, the state reached after reading $uv^nv[1{\colon}l]$  in the run $\rho$ is $q_1$ for infinitely many $n$s. Let $v'=v[l+1\colon |v|]v[1\colon l]$, let $u'=uv[1{\colon}l]$ and let $w'=v[l+1\colon |v|]$, we have that $\tuple{\tuple{u'v'^nw'z^\omega}_{n\in \Nn},\tuple{u'v'^\omega}_{n\in \Nn}}$ is a bad pair.
	To simplify notations, we write this pair again as $\tuple{\tuple{uv^nwz^\omega}_{n\in \Nn},\tuple{uv^\omega}_{n\in \Nn}}$.
	We have gained that on the run over $uv^\omega$, the accepting state $q_1$ is reached infinitely often after reading $v$ factors. Now let $k,l$ be integers such that ${p_1}\xrightarrow{uv^k} q_1\xrightarrow {v^l}q_1$, with $p_1$ an initial state.
	We consider $l$ different sequences, for $r\in \set{0,\ldots,l-1}$ we define the sequence $s_r=\tuple{uv^{ln+r}wz^\omega}_{n\in \Nn}$. Since $\tuple{\tuple{uv^nwz^\omega}_{n\in \Nn},\tuple{uv^\omega}_{n\in \Nn}}$ is a bad pair, there must be a value $r$ such that $\tuple{\tuple{(uv^k)(v^l)^n(v^rw)z^\omega}_{n\in \Nn},\tuple{(uv^k)(v^l)^\omega}_{n\in \Nn}}$ is a bad pair.
	To simplify notations, again, we rename this pair $\tuple{\tuple{uv^nwz^\omega}_{n\in \Nn},\tuple{uv^\omega}_{n\in \Nn}}$, and we know that ${p_1}\xrightarrow{u} q_1\xrightarrow {v}q_1$, with $p_1$ initial and $q_1$ accepting.

	For any $m$ larger than the number of states $|Q|$ of $T$,  let us consider the run of $T$ over $uv^mwz^\omega$.
	Let $i_m,j_m, k_m$ be such that we have the accepting run ${p_m}\xrightarrow{uv^{i_m}} q_m\xrightarrow {v^{j_m}}q_m\xrightarrow{v^{k_m}wz^\omega}$, with $i_m+j_m+k_m=m$ and $0<j_m\leq |Q|$. And let $r_m=i_m+k_m \mod j_m$.
	Let us consider for some $m$, the sequence $\tuple{(uv^{r_m}) (v^{j_m})^n (wz)^\omega}_{n\in \Nn}$.
	By the presence of these loops, each of these sequences has a convergent image.
	There is actually a finite number of such sequences, since $j_m$ and $r_m$ take bounded values.
	Since the original pair is bad, this means that there must exist $m$ such that the sequence $\tuple{f((uv^{r_m}) (v^{j_m})^n (wz)^\omega)}_{n\in \Nn}$ does not converge to $f(uv^\omega)$.
	Let $u'=uv^{i_m}$, $v'=v^{j_m}$ and $w'=v^{k_m}w$, then
	$\tuple{\tuple{u'v'^nw'z^\omega}_{n\in \Nn},\tuple{u'v'^\omega}_{n\in
			\Nn}}$ is a bad pair. Once again we rename the pair
	$\tuple{\tuple{uv^nwz^\omega}_{n\in \Nn},\tuple{uv^\omega}_{n\in
			\Nn}}$ and now we have both ${p_1}\xrightarrow{u} q_1\xrightarrow
	{v}q_1$ and ${p_2}\xrightarrow{u} q_2\xrightarrow {v}q_2$, such that
	$p_1,p_2$ are initial, $q_1$ is accepting and $wz^\omega$ has a final run from $q_2$.
	
	Now we have established the shape of the pattern, we only have left to show the mismatch properties.
	$$\begin{array}{lll}
		p_1 \xrightarrow{u|u_1} q_1 \xrightarrow{v|v_1} q_1\\
		p_2 \xrightarrow{u|u_2} q_2 \xrightarrow{v|v_2} q_2\xrightarrow{wz^\omega |y_2}\\
	\end{array}$$
	
	Let us first assume that $v_2\neq \epsilon$. Then $\lim_nf(uv^nwz^\omega)=u_2v_2^\omega$.
	Since the pair is bad, there exists $k$ such that $\mismatch(u_1v_1^k, u_2v_2^k)$. Hence, up to taking $u'=uv^k$, we have established the pattern:
	$$\begin{array}{lll}
		\exists \pi_1\colon p_1 \xrightarrow{u'|u_1'} q_1,\ \exists \pi_1'\colon q_1 \xrightarrow{v|v_1} q_1\\
		\exists \pi_2\colon p_2 \xrightarrow{u'|u_2'} q_2,\ \exists \pi_2'\colon q_2 \xrightarrow{v|v_2} q_2\\
		\big(\mathsf{init}(p_1)\wedge \mathsf{init}(p_2)\wedge\mathsf{acc}(q_1)\big)\wedge
		\big( \mismatch(u_1',u_2') \big)
	\end{array}
	$$
	
	Let us now assume that $v_2= \epsilon$. Then $\lim_nf(uv^nwz^\omega)=u_2y_2$.
	Then there is  a prefix $w_2$ of $y_2$ and an integer $k$ such that $\mismatch(u_1v_1^k, u_2w_2)$.
	Hence, up to taking $u'=uv^k$, and $w'$ a sufficiently long prefix of $wz^\omega $ we have established the pattern:
	
	$$\begin{array}{lll}
		\exists \pi_1\colon p_1 \xrightarrow{u|u_1} q_1,\ \exists \pi_1'\colon q_1 \xrightarrow{v|v_1} q_1\\
		\exists \pi_2\colon p_2 \xrightarrow{u|u_2} q_2,\ \exists \pi_2'\colon q_2 \xrightarrow{v|v_2} q_2,\ \exists \pi_2''\colon q_2 \xrightarrow{w|w_2} r_2\\
		\big(\mathsf{init}(p_1)\wedge \mathsf{init}(p_2)\wedge\mathsf{acc}(q_1)\big)\wedge
		\big( (v_2=\epsilon\wedge \mismatch(u_1,u_2w_2)) \big)
	\end{array}
	$$
	which concludes the proof.
\end{proof}

\begin{thm}\label{thm:decrat}
	Deciding if a one way B\"uchi transducer defines a continuous
	(resp. uniformly continuous) function can be done in $\nlogspace$.
\end{thm}
\begin{proof}
	Let $T$ be a one way B\"uchi transducer defining a function $f$. From
	Lemma~\ref{lem:pattern-rat}, if $T$ is trim, non-continuity of $f$
	is equivalent to $T$ satisfying the formula $\phi_{cont}$ of
	Figure~\ref{fig:pattern-1}. This formula is expressed in the syntax of
	the pattern logic from \cite{DBLP:conf/dlt/FiliotMR18}, where  
	it is proved that
	model-checking pattern formulas against transducers can be done in
	$\nlogspace$~\cite[Thm.~6]{DBLP:conf/dlt/FiliotMR18}. This yields the result.

	If $T$ is not trim, then we modify the formula $\phi_{cont}$ to
	additionally express that there must be some accepting run from
	$r_2$ on some input. Equivalently, we express that there exists a
	run from $r_2$ to some accepting state $s$, and a run looping in $s$,
	in the following way: we just add the quantifiers $\exists
	\pi_3 \colon r_2\xrightarrow{\alpha\mid \beta} s\ \exists
	\pi_4\colon s\xrightarrow{\gamma\mid \tau} s$ to $\phi_{cont}$ and the
	constraint $\textsf{acc}(s)$ which requires $s$ to be accepting. 
	
	The proof  for non-uniform continuity, using formula
	$\phi_{\text{u-cont}}$ is similar. 
\end{proof}

\subsection{Regular case} \label{subsec:reg}
As for the rational case, we first define a
pattern which is satisfied by a \twoDFTpla if and only if it defines a
non-continuous function. Then we show how to check for the existence of
this pattern. It is however more intricate than in the rational
setting. First, we get rid of the look-ahead by working on words
annotated with look-ahead information. We now make this notion
formal.

\subsubsection{Look-ahead removal and annotated words}\label{subsubsec:la} Let $f$ be realized by a \twoDFTpla $\Tt$ over alphabet $\Sigma$ with look-ahead
automaton $P$. Let $Q_P$ be the states of
$P$. We define $\widetilde{\Tt}$, a
\twoDBT over $\Sigma\times Q_P$ which, informally, simulates $\Tt$ over words annotated
with look-ahead states, and which accepts only words with a correct
look-ahead annotation with respect to $P$. Formally, for an annotated
word $u\in (\Sigma\times Q_P)^\infty$, we denote by $\pi_\Sigma(u) \in
\Sigma^\infty$ its projection on $\Sigma$. An annotated word $x\in
(\Sigma\times Q_P)^\omega$ is \emph{well-labelled} if $\pi_{Q_P}(u)$ is the accepting run of $P$ over $u$.

As $P$ is prophetic, for all words
$u\in\dom{f}$, there exists a unique well-annotated word, denoted
$\tilde{u}$, which satisfies $\pi_\Sigma(\tilde u)=u$. We now let
$\tilde{f}$ be the function defined, for all $u\in\dom{f}$, by
$\tilde{f}(\tilde{u}) = f(u)$. The function $\tilde{f}$ is easily seen
to be definable by a \twoDBT which simulates $\Tt$ over
the $\Sigma$-component of the input word and simulates $P$ over the
$Q_P$-component of the input word to check that the annotation is
correct. 

\begin{prop}
	\label{prop:criticalpatterntwoway}
    One can construct, in linear-time, a \twoDBT denoted
    $\widetilde{T}$ which realises the function $\tilde{f}$. 
\end{prop}

\subsubsection{Transition monoid of \twoDBT} In our pattern, we will
be using the notion of transition monoid of a \twoDBT, as the
transition monoid of its underlying two-way automaton (obtained by
ignoring the output words on transitions).

Let $T$ be a \twoDBT, we define its transition monoid $M_T$ in the
usual way (see \eg\cite{DBLP:journals/lmcs/BaschenisGMP18} for
details), but we also add information telling us if a word produces
empty output or not. The set $M_T$ contains functions
$Q\times\set{L,R} \rightarrow Q\times
\set{L,R}\times\set{0,1}$. Informally, the transition morphism of $T$,
$\phi_T\colon \Sigma^*\rightarrow M_T$ sends a word $u$ to the function
mapping \eg$(p,L)$ to $(q,R,1)$ if there exists a run of
$T$ on $u$ which enters $u$ from the left in state $p$, stays within
the inner positions of $u$ until it exits $u$ from the right in state
$q$, and $T$ has produced a
non-empty output. This run is called a left-to-right traversal
(LR-traversal for short). Left-to-left (LL-), right-to-right (RR-) and
right-to-left (RL-) traversals are defined similarly. For instance a
run from $p$ to $q$ is an LL-traversal on $u$, if it enters $u$ from
the left in state $p$, stays within the inner positions of $u$, until
it exists $u$ from the left in state $q$. If this run did not produce
any output, then $\phi_T(u)$ contains the function mapping
$(p,L)$ to $(q,L,0)$.

The transition monoid of a two-way automaton is defined similarly,
except that the information on whether there is a production of
non-empty output or not is not required, \ie it is a set of
functions of type $Q\times\set{L,R} \rightarrow Q\times
\set{L,R}$.

\subsubsection{Critical Pattern}\label{subsec:crit-pat}
 Let $\Tt$ be a \twoDFTpla and
$\widetilde{T}$ be its associated \twoDBT running over annotated words
(see Section~\ref{subsubsec:la}). 
Let $\rho=(q_0,i_0=0) \stackrel{b_{i_0}/\gamma_1}{\longrightarrow}
(q_1, i_1)\stackrel{b_{i_1}/\gamma_2}{\longrightarrow}\cdots
(q_n,i_n=l)$ be an initial run of $\widetilde{T}$ over a finite word $uvw$.
We say that the factor $v$ is \emph{non-producing} in $\rho$ if 
all the 
transitions occurring within $v$ output $\epsilon$. Otherwise $v$ is \emph{producing}.
Formally, if in all the transitions $(q_{j},i_j) \stackrel{b_{i_j}/\gamma_j}{\longrightarrow} (q_{j+1}, i_{j+1})$ in run $\rho$ such that 
$|u|\leq i_j<|uv|$ and $\gamma_j=\epsilon$, then $v$ is non-producing.

We define the notion of \emph{critical pattern} for the \twoDBT
$\widetilde{T}$. We say that $\widetilde{T}$ admits 
a critical pattern if there are 
words $u,v\in\Sigma^*$, $u_1,v_1,w_1,z_1,u_2,v_2,w_2,z_2\in (\Sigma\times Q_P)^*$ such
that
\begin{enumerate}
  \item $\pi_\Sigma(u_1) = \pi_\Sigma(u_2) = u$, $\pi_\Sigma(v_1) =
    \pi_\Sigma(v_2) = v$,
  \item $v_1$ and $v_2$ are idempotent\footnote{$\phi_{\widetilde T}(v_i)=\phi_{\widetilde T}(v_iv_i)$ for $i\in \set{1,2}$} with respect
    to the transition monoid of $\widetilde{T}$,
    \item there exist two accepting runs of $\widetilde{T}$ on
      $u_1v_1w_1z_1^\omega$ and $u_2v_2w_2z_2^\omega$ which can be
      respectively decomposed into $\rho_1\lambda_1$ and
      $\rho_2\lambda_2$ such that for all $i=1,2$,
    \item $\rho_i$ is over $u_iv_iw_i$ and does not end in $v_i$,
    \item $v_i$ is not producing in $\rho_i$,
    \item there is a mismatch between $\out{\rho_1}$ and $\out{\rho_2}$.
\end{enumerate}

We give the following characterisation of continuity and uniform continuity. Note that the patterns are similar to the one-way case.
\begin{prop} \label{prop:cont-crit-pat}
	Let $T$ be a \twoDFTpla realising a function $f$.
	Then the following holds:
\begin{enumerate}
	\item $f$ is not uniformly continuous (resp.~continuous).
	\item the \twoDBT $\tilde{T}$ admits a critical pattern $u_1,u_2,v_1,v_2,w_1,z_1,w_2,z_2,\rho_1,\lambda_1,$ $\rho_2,\lambda_2$ (resp.~and $\pi_\Sigma(u_1v_1^\omega)\in \dom f$)
\end{enumerate}
\end{prop}
\begin{proof}
	We will show that the existence of a critical pattern is equivalent to the existence of a synchronised bad pair. From Proposition~\ref{prop:bad-pairs} this will conclude the proof.
	
	Let us first assume that we have a critical pattern $u_1,u_2,v_1,v_2,w_1,z_1,w_2,z_2,\rho_1,\lambda_1,\rho_2,\lambda_2$.
	Our goal is to prove that $\tuple{\tuple{\pi_\Sigma(u_1v_1^nw_1z_1^\omega)}_n,\tuple{\pi_\Sigma(u_2v_2^nw_2z_2^\omega)}_n}$ is a bad pair.	
	 We define $\rho_1^n, \rho_2^n$ the runs obtained from $\rho_1,\rho_2$, just by iterating the respective $v_1,v_2$ factors $n$ times. Since $v_1,v_2$ are idempotents, reading $v_i$ or $v_i^n$ does not change the shape of the run $\rho_i$, nor the order of the outputs occurring outside of $v_i$.
	Moreover, since $v_i$ is non-producing in both runs, the outputs over $v_i$ and $v_i^n$ are also the same. Hence we get $\out {\rho_1}=\out {\rho_1^n}$ and $\out{\rho_2}=\out{\rho_2^n}$. Thus the runs $\rho_1^n\lambda_1^n$ and $\rho_2^n\lambda_2^n$ produce outputs which mismatch at a position smaller than $\max(|\rho_1|,|\rho_2|)$ and the pair $\tuple{\tuple{\pi_\Sigma(u_1v_1^nw_1z_1^\omega)}_n,\tuple{\pi_\Sigma(u_2v_2^nw_2z_2^\omega)}_n}$ is indeed bad.

	Conversely, let us assume that $\tuple{\tuple{uv^nx_1y_1^\omega}_n,\tuple{uv^nx_2y_2^\omega}_n}$ is a bad pair.
	Up to extracting subsequences, we can assume that $v,y_1,y_2$ are idempotent both for $T$ and its look-ahead automaton.
	Thus we can obtain, up to taking a larger word $u$, a synchronised bad pair of the form $\tuple{\tuple{\pi_\Sigma(u_1v_1^nw_1z_1^\omega)}_n,\tuple{\pi_\Sigma(u_2v_2^nw_2z_2^\omega)}_n}$ with $\pi_\Sigma(u_1)=\pi_\Sigma(u_2)=u$ and $\pi_\Sigma(v_1)=\pi_\Sigma(v_2)=v$. We also write $\pi_\Sigma(w_i)=x_i$ and $\pi_\Sigma(z_i)=y_i$ for $i\in \set{1,2}$.
	Since the pair is bad there exists $i$ such that for any $m$, there exists $n\geq m$ with $|f(uv^nx_1y_1^\omega)\wedge f(uv^nx_2y_2^\omega) |\leq i$.
	Let $n$ be larger than $2i+2$ with $|f(uv^nx_1y_1^\omega)\wedge f(uv^nx_2y_2^\omega) |= j\leq i$.
	Let $\rho_1$ (resp.~$\rho_2$) be the prefix of the run over $u_1v_1^nw_1z_1^\omega$ (resp.~$u_2v_2^nw_2z_2^\omega$) which stops after producing the $j$th output.
	We make the following claim with respect to the sequence $u_1v_1^{i+1}v_1^kv_1^{i+1}w_1z_1^\omega$:
	
	\begin{claim}
		\label{claim:output-pattern}
	Any transition of $\rho_1$ occurring within the factor $v_1^k$ must produce empty outputs and $\rho_1$ ends outside of this factor.
	\end{claim}
	
	If we assume that the claim is true, then it is similarly true for $\rho_2$. Then we obtain a critical pattern $u_1v_1^{i+1},u_2v_2^{i+1},v_1^k,v_2^k,v_1^{i+1}w_1,z_1,v_1^{i+1}w_2,z_2,\rho_1,\lambda_1,\rho_2,\lambda_2$.
	
	We only have left to show that the claim holds.
	\begin{proof}[Proof of Claim~\ref{claim:output-pattern}]
		Let us assume towards a contradiction that a transition occurring within $v_1^k$ in $\rho_1$ has non-empty output. Let $\rho_1'$ the shortest prefix of the run $\lambda$ over $u_1v_1^{i+1}v_1^kv_1^{i+1}w_1z1^\omega$ which extends $\rho_1$ and ends outside of the $v_1^k$ factor. Note that if $\rho_1$ already ends outside of this factor, $\rho_1'=\rho_1$.
		Let $\rho$ be a factor of $\rho_1'$ which contains such a producing transition and which starts and ends at a boundary of the $v_1^k$ factor without leaving it.

		Since $\tilde T$ is deterministic and since $v_1$ is idempotent, all $v_1$ factors (except possibly for the first and last ones) must also produce such a non-empty output in $\lambda$.
		These outputs must occur in $\lambda$ either in increasing or decreasing order with respect to the order of the input word. If they occur in increasing order, the first ${i+1}$ $v_1$ factors produce non-empty outputs within $\rho_1$, which contradicts the assumption that $|\out{\rho_1}| \leq i$. Similarly if the outputs occur in decreasing order then the last ${i+1}$ $v_1$ factors produce non-empty outputs within $\rho_1$, which also yields a contradiction, hence we obtain that the claim must hold.
\qedhere
	\end{proof}
    \noindent
    This proves \autoref{prop:cont-crit-pat}.
\qedhere	
\end{proof}

\subsubsection{Deciding uniform continuity and continuity for regular
  functions}

We now show how to decide, in \textsc{PSpace}, whether a given
\twoDFTpla defines a uniform continuous  function
by checking whether its associated \twoDBT over annoted words
admits a critical pattern. Deciding continuity is based on the same
ideas and shown in the last paragraph of this section. 
We reduce the problem to
checking the emptiness of a non-deterministic two-way Parikh
automaton~\cite{DBLP:conf/dcfs/Ibarra14}. A \emph{two-way Parikh
automaton} (\twoPA) $\mathcal{P}$ of dimension $d$ is a two-way automaton, running on \emph{finite} words, extended with vectors
of integers (including $0$) 
of dimension $d$. Each
dimension can be seen as a counter which can only be incremented. 
A run on a finite word is accepting if it reaches some
accepting state and the (pairwise) sum of the vectors met along the transitions
belong to some given semi-linear set represented as some
existential Presburger formula\footnote{We recall that existential
  Presburger formulas are existentially quantified first-order
  formulas over the predicate $t_1\leq t_2$ where $t_i$ are terms over 
  the constants $0,1$ and the addition $+$.}. The emptiness
problem for \twoPA is undecidable but decidable in \textsc{PSpace} for
\emph{bounded-visit} \twoPA~\cite{DBLP:conf/fsttcs/FiliotGM19}. A
\twoPA is bounded visit if there exists some $k$ such that in any accepting run of
$\mathcal{P}$, any input position is visited at most $k$ times by that
run. The \twoPA we construct will be of polynomial size and
bounded-visit, entailing \textsc{PSpace} decidability of uniform continuity for \twoDFTpla, following
Proposition~\ref{prop:criticalpatterntwoway}.

\begin{lem}\label{lem:decpat}
    Given a \twoDFTpla $T$, one can construct in polynomial time a bounded-visit \twoPA
    $\mathcal{P}_T$ (of polynomial size) such that
    $L(\mathcal{P}_T)\neq\emptyset$ if and only if $\widetilde{T}$ admits a critical
    pattern. 
\end{lem}

Before proving the lemma, let us state our main result.

\begin{thm}\label{thm:main1}
    Deciding whether a regular function given as a deterministic
    two-way transducer with prophetic B\"uchi look-ahead is
    uniformly continuous is \textsc{PSpace-Complete}. Hardness also holds for
    continuity. 
\end{thm}

\begin{proof}
    The upper-bound is a direct consequence of
    Proposition~\ref{prop:criticalpatterntwoway},
    Lemma~\ref{lem:decpat}, and the \textsc{PSpace} upper-bound
    of~\cite{DBLP:conf/fsttcs/FiliotGM19} for the non-emptiness of
    bounded-visit two-way Parikh automata. 

    The lower-bound is easily obtained by reducing the problem of
    checking the non-emptiness of the language of a deterministic
    two-way automaton over finite words, which is known to be
    \textsc{PSpace}-hard\footnote{This is a folklore result, easily
      obtained by reducing the \textsc{PSpace}-hard problem of checking the non-emptiness
      of the intersection of the languages of $n$ DFA, each pass of
      the two-way automaton over the whole input simulates one pass of
      each DFA.}. Given a deterministic two-way automaton $A$ on
    finite words over some alphabet $\Sigma$ containing at least two
    letters $a$ and $b$, we construct a \twoDFTpla $T$ defining the following
    function $f$. It is defined on all $\omega$-words $u$ in
    $(\Sigma\cup\{\#\})^\omega$ by $f(u) = a^\omega$ if $u = v\# w$
    for some $v\in L(A)$ and $w$ contains infinitely many $a$,
    otherwise by $f(u) = b^\omega$. Clearly, if $L(A)$ is empty, then
    $f$ is the constant function mapping any word to $b^\omega$ and is
    therefore uniformly continuous (and continuous). Conversely, if there
    exists some $v\in L(A)$, then we claim that $f$ is not
    continuous (and hence not uniformly continuous). Indeed, consider the family of words $u_n =
    v\#a^n\#^\omega$ for all $n\geq 0$ and some $a\in\Sigma$. Clearly,
    $f(u_n) = b^\omega$ for all $n$ as $u_n$ contains finitely many
    $a$. However
    $f(\lim_{n\rightarrow\infty} u_n) = f(v\#a^\omega) = a^\omega$, which is
    different from $\lim_{n\rightarrow\infty} f(u_n) = b^\omega$. Finally, it
    remains to show how to construct $T$ from $A$. The idea is to use
    a look-ahead to check whether the input is of the form $v\# w$
    where $v \in \Sigma^*$ and $w$ contains infinitely many $a$. 
    If that is the case, then $T$ simulates $A$ on $v$ (using $\#$ as an
    endmarker). At the end of this simulation, if $A$ reaches an accepting state, then $T$
    reads $w$ and keeps on producing $a$.
    On the other hand, if $A$ reaches a non-accepting state, then $T$ keeps on producing $b$ on the $w$ part.  If the look-ahead has failed (meaning that the input is not
    of the above form), then $T$ keeps on producing $b$ while reading
    its input in a one-way fashion. 
\end{proof}

We now proceed to the proof of Lemma~\ref{lem:decpat}. 

\paragraph{Encoding of the set of critical patterns as a language} We show how to
encode the set of critical patterns as the language of a \twoPA. To
that end, let us see how a critical pattern can be encoded as a word. 
The \twoPA will have to check properties $(1)-(6)$ of the definition of
a critical pattern. The first condition is about projections. To check
that $u_1,u_2$ project on the same $\Sigma$-word, as well as
$v_1,v_2$, we just encode them as words over the alphabet
$\Sigma\times Q_P\times Q_P$. Let us make this more formal.
Let $\#$ be a fresh symbol.     For any two alphabets $A,B$ and any two words $w_1\in A^*$ and
    $w_2\in B^*$ of same length, we let $w_1\otimes w_2\in (A\times
    B)^*$ be their convolution, defined by $(w_1\otimes w_2)[i] =
    (w_1[i],w_2[i])$ for any position $i$ of $w_1$.

    We now define a language $\text{Crit}_{\Tt} ~{\subseteq}~\vdash ((\Sigma\times Q_P^2)^*\#)^2((\Sigma\times Q_P)^*\#)^4\dashv$
    which consists of words of the form 
    $w ~{=}~ \vdash (u \otimes a_1 \otimes a_2)\# (v \otimes b_1\otimes b_2)\#
    w_1\# z_1\# w_2\# z_2\#\dashv$ where, if $u_i = u\otimes a_i$ and $v_i =
    u\otimes b_i$, then $u_1,v_1,w_1,z_1,u_2,v_2,w_2,z_2$ is a
    critical pattern. 

    The following proposition is immediate
    \begin{prop}\label{prop:trivial}
        $\Tt$ admits a critical pattern if and only if $\text{Crit}_\Tt\neq\emptyset$.
    \end{prop}

    Note that this encoding makes sure condition $(1)$ of the
    definition of critical pattern is ensured, but the other
    properties will have to be checked by the $\twoPA$. 

    \paragraph{Checking idempotency with a succinct two-way automaton}
    The \twoPA has to check condition $(2)$ of the critical patterns,
    \ie idempotency with respect the transition monoid of the
    \twoDBT $\widetilde{T}$. We show more generally, given a
    deterministic two-way automaton $A$, how a two-way automaton of
    polynomial size can
    recognise idempotent words with respect to the transition monoid
    of $A$. The proof of this result is not needed to understand the proof
of the main lemma (Lemma~\ref{lem:critpa}), namely that the set $\text{Crit}_\Tt$ is recognisable by a
\twoPA, and can be skipped in a first reading.

    \begin{lem}\label{lem:idempotentregular}
        For any deterministic two-way automaton $A$
        (resp. deterministic two-way B\"uchi transducer $T$), one can construct
        in polynomial time a (bounded-visit) deterministic two-way automaton $I_A$
        which recognises the set of words $\# u \#$ such that $u$ is idempotent with respect
        to $A$, and $\#$ is a fresh symbol.  
    \end{lem}

    \begin{proof}
        We prove the result for automata. Its adaptation to
        transducers is trivial, as the transition monoid of a
        transducer is the same as the transition monoid of its
        underlying automaton obtained by ignoring the outputs, without
        the information on whether some non-empty
        output is produced or not by a traversal. The following
        construction, done for automata, can be trivially adapted to
        account for this information.

        Let $Q$ be the set of states of $A$. Without loss of generality, we can assume that
        $Q = \{1,\dots,n\}$. Let $(i,j)\in Q^2$. We explain how a
        two-way automaton $A_{i,j}^{LR}$ can check whether $(i,j)$ is
        is left-to-right traversal of $u$: it needs to start from the
        left of $u$ in state $i$, and simulates $A$ until it reaches a
        boundary of $u$: if it is the left boundary, it rejects, if it
        the right boundary, it accepts. Now, let us explain how a
        two-way automaton $B_{i,j}^{L,R}$ can check whether $(i,j)$ is
        a left-to-right traversal of $A$ over $uu$, by reading only
        $\#u\#$. Intuitively, the main idea is to simulate $A$ and
        when $A$ wants to go to the right of $u$ following some
        transition $t$, then $B_{i,j}^{L,R}$
        comes back to the left of $u$ and applies $t$ and proceeds
        with its simulation of $A$. More precisely, $B_{i,j}^{L,R}$
        has two modes: $M_1$ and $M_2$. It stays in mode $M_1$ as long
        as $A$ does not cross the right boundary of $u$ or crosses the
        left boundary (in which case it stops its computation in a
        non-accepting state). As soon as it
        crosses the right boundary, using some transition $t$, $B_{i,j}^{L,R}$ moves to mode
        $M_2$, comes back to the left of $u$, and applies $t$. It then
        simulates $A$ until it reaches the right boundary in state $j$
        in which case it stops its computation and accepts, or it
        reaches the left boundary in which case it stops its
        computation and rejects.

        Note that in the above constructions, $A_{i,j}^{LR}$ and
        $B_{i,j}^{LR}$ are both deterministic, since $A$ is
        deterministic. Therefore, they have bounded
        crossing. Moreover, they have a polynomial number of states. We construct in a similar
        way automata for each type of traversals (LR, RL, LL, RR) and
        each pair $(i,j)$.

        Then, clearly, $u$ is idempotent if and only if the following holds
        $$
        \bigwedge_{(i,j)\in Q^2, \text{trav}\in\{LL,RL,LR,RR\}} \#u\#\in
        L(A_{i,j}^{\text{trav}})\Leftrightarrow  \# u \#\in L(B_{i,j}^{\text{trav}})
        $$

        Since the automata are deterministic, they can be complemented
        in polynomial time. Moreover, intersection of two-way automata
        can be done in linear time by serial composition: the first
        automaton is simulated and if it eventually accepts, then the
        other automaton is simulated and must accept as well.  
        Using these results, one can construct in polynomial time a
        deterministic two-way automaton implementing the conjunction
        above, concluding the proof. 
    \end{proof}

\paragraph{Checking acceptance of lasso words by a succinct two-way
  automaton over finite words} 
We prove a useful technical lemma, which intuitively says that it is possible to construct a two-way
automaton of polynomial size accepting words of the form $u\#v$ such
that $uv^\omega$ is accepted by a given B\"uchi two-way
automaton $A$ and $v$ is idempotent with respect to the transition
monoid of $A$. The statement of the lemma is slightly more
general. The proof of this lemma is not needed to understand the proof
of the main lemma, namely that the set $\text{Crit}_\Tt$ is recognisable by a
\twoPA, and can be skipped in a first reading.

A \emph{marked word} over an alphabet $\Sigma$ is an $\omega$-word
over $\Sigma\times \{0,1\}$ which contains exactly one occurrence of
$1$. Given $i\geq 1$ and $x\in\Sigma^\omega$, we denote by $x\triangleright
i$ the marked word such that for all $j\geq 1$, $(x\triangleright i)[j] =
(x[j],0)$ if $j\neq i$ and $(x\triangleright i)[j] = (x[j],1)$ otherwise.

\begin{lem}\label{lem:lasso}
    Let $A$ be a non-deterministic (bounded-visit) two-way automaton over an alphabet
    $\Sigma$, $\#\not\in\Sigma$ and $q$ be a state of $A$. Let
    $L_q$ be the language of words $({\vdash} u\#v{\dashv})\triangleright i$ such that
    $u,v\in\Sigma^*$, $2\leq i\leq |u|+1$, $v$ is idempotent with
    respect to the transition monoid of $A$,  and there exists an accepting run of $A$ on
    $uv^\omega$ starting in configuration $(q,i)$. Then, $L_q$ is recognisable by a
    non-deterministic bounded-visit two-way automaton of polynomial
    size (in $A$).
\end{lem}

\begin{proof}
    We define a two-way automaton $P$
    recognising $L_q$. The symbol $\#$
    is needed to identify the left boundary of $v$.
    First, $P$ checks that the input is valid, \ie it is of the form 
    $({\vdash} u\#v{\dashv})\triangleright i$ such that 
    $2\leq i\leq |u|+1$ and $v$ is idempotent with
    respect to the transition monoid of $A$. To check idempotency, we
    rely on the construction of Lemma~\ref{lem:idempotentregular}.

    If the validity checking succeeds, $P$ comes back to the initial
    position of its input and check the existence of an accepting run
    as required by the statement of the lemma. Since this accepting
    run is on $uv^\omega$ and $P$ can only read one occurrence of $v$,
    we first give a characterisation of such an accepting run which
    will turn useful to check its existence while accessing only one
    iteration of $v$. This characterisation  intuitively says that there is an
    accepting loop on some finite iteration $v^\kappa$ accessible from
    the marked position $i$ in state
    $q$. More precisely, it can be seen that any accepting run $r$ of
    $A$ on $uv^\omega$ starting in configuration $(q,i)$ can be
    decomposed as:
    $$
    r = \rho_0L_1R_1\dots L_nR_n\rho_1\ell^\omega
    $$
    where:
    \begin{enumerate}
      \item $n$ is smaller than the number of states of $A$
      \item $\rho_0$ is a run on $u$ from configuration $(q,i)$ to some
        configuration $(q_1, |u|+1)$
      \item for all $1\leq j\leq n$, $L_j$ is a left-to-left traversal on
        $v^\omega$ starting in some configuration $(q_j,|u|+1)$ and
        ending in some configuration $(p_j,|u|)$
      \item for all $1\leq j\leq n$, $R_j$ is a right-to-right traversal on
        $u$ starting in configuration $(p_j,|u|)$ and
        ending in some configuration $(q_j,|u|+1)$
      \item $\rho_1$ is a finite run visiting only positions of $v^\omega$ from configuration $(q_n,
        |u|+1)$ to some configuration $(p, j_1)$ such that $j_1 =
        |u|+|v|.m+1$ for some $m$ (\ie $j_1$ is the initial position
        of some iteration of $v$) 
      \item $\ell$ is a finite run visiting only positions of
        $v^\omega$  from configuration $(p,j_1)$
        to some configuration $(p,j_2)$ such that $j_2 = j_1+|v|m$ for
        some $m\geq 1$ (\ie$j_2$ is the initial position of some
        iteration of $v$, and is at the right of $j_1$)
    \end{enumerate}

    In the above characterisation, note that $L_j,R_j$ are either LL-
    or RR-traversal starting at the right boundary of $u$. 
    Since $r$ is accepting, it visits the whole infinite input $uv^\omega$ and
	therefore there cannot be more such traversals than the number of states. Otherwise,
	this boundary would be visited twice in the same state by $A$, and
	therefore, it would be visited infinitely many times,
	contradicting that the whole input is read. 
	This justifies why we
    can take $n$ smaller than the number of states.

    This characterisation is not directly exploitable by $P$. We make
    the following observation: since $v$ is idempotent, $L_j$ is a
    left-to-left traversal on $vv$ for all $1\leq j\leq n$. Indeed,
    suppose that is not the case, it means that
    $L_j$ visits all positions of $v^2\sigma$ at least once, where
    $\sigma$ is the first symbol of $v$. Let $s_1$ be the state
    of $L_j$ the first time it reaches the end of $v$, and $s_2$ be
    the state of $L_j$ the first time it reaches the end of $vv$. In
    other words, there is a left-to-right traversal from $q_j$ to
    $s_1$ on $v$ and a left-to-right traversal from $q_j$ to $s_2$ on
    $vv$. Since $v$ is idempotent, we get that $s_1=s_2$. This
    contradicts that $L_j$ is a left-to-right traversal, since $A$
    then loops on $s_1=s_2$ and  moves forever to the right when
    initially starting to read $v^\omega$ on state $q_j$. This means
    that to check the existence of a left-to-left traversal on
    $v^\omega$, one only needs to consider two iterations of $v$.

    We now proceed to the construction of $P$.  The automaton $P$
    guesses a decomposition as in the above characterisation by simulating $A$. The main
    difficulty is that it reads only one occurrence of $v$ while $A$
    must be simulated on $uv^\omega$. Let us first explain how the
    decomposition can be guessed. First, $P$ accesses the marked
    position $i$ and start simulating $A$ from state $q$ (ignoring the
    marking in $\{0,1\}$ since $A$ is over alphabet $\Sigma$). If
    eventually $A$ reaches the right boundary of $u$ in some state
    $q_1$, $P$ guesses some state $p_1$ and checks whether there exists
    a left-to-left traversal $L_1$ of $A$, from $q_1$ to some $p_1$, on
    $v^\omega$ (we explain later how this verification can be done by
    $P$). If it is the case, $P$ again simulates $A$ on $u$ and if
    eventually $A$ reaches the right boundary of $u$ in some state
    $q_2$, once again, $P$ checks whether there exists a left-to-left
    traversal of $v^\omega$ from $q_2$ to some $p_2$, and so on for at
    most $n$ times. Non-deterministically, when $P$ reaches the right
    boundary of $u$, it can decide to check for the existence of an
    accessible accepting loop in $v^\omega$, \ie runs $\rho_1$ and
    $\ell$ as in the above characterisation (we also explain later how this
    verification can be done by $P$). If such a loop is found, $P$
    accepts.

    Let us now explain how the two verifications involved in the
    latter simulation can be done by $P$, namely: checking the
    existence of a left-to-left traversal on $v^\omega$ from some
    state, say $s$, and checking the existence of an accessible accepting
    loop. The main difficulty is that $P$ has access to only one
    occurrence of $v$, but it can use two-wayness to overcome this
    problem. If $A$ needs to cross the right
    boundary of $v$ from the left (\ie access the next iteration of $v$), $P$ just
    rewinds its reading head to the first position of $v$, and
    if $A$ needs to cross the left boundary of $v$ from the right
    (\ie access the previous iteration of $v$), then $P$ moves
    its reading head to the last position of $v$. We call this
    simulation a \emph{simulation modulo}. We use this idea of
    simulation modulo to check the existence of LL-traversals as well
    as $\rho_1$ and $\ell$.

    For LL-traversals, as explained before,
    one only needs to check the existence of an LL-traversal on
    $vv$. $P$ therefore performs a simulation of $A$ modulo and uses a
    bit of information to check whether it is in the first or second
    occurrence of $v$. 

    To check the existence of $\rho_1$, we can use similar arguments
    as before, based on the fact that $v$ is idempotent, to prove that
    $\rho_1$ only visits the two first iterations of $v$. This is due
    to the fact that $\rho_1$ can be decomposed into a sequence of
    traversals on some powers of $v$ but each of those
    traversal is a traversal on $v$ by idempotency. For example,
    $\rho_1$ might be an LR-traversal on $v^{\kappa_1}$ followed by a
    LL-traversal on $v^{\kappa_2}$, followed by a RR-traversal on $v^{\kappa_3}$, up to
    reaching configuration $(p,j_1)$. However by idempotency, each of
    these traversals are actually traversals on $v$, so, $\kappa_j =
    1$ for all $j$. Hence, only two iterations of $v$ are needed (two
    and not one, because, for instance, an LR-traversal on $v$ can be
    followed by a LL-traversal on $v$, and hence two iterations are
    needed). Since two iterations of $v$ are needed, $P$ can check the
    existence of $\rho_1$ using some simulation modulo and one bit of
    information.

    Finally, it remains for $P$ to check for the existence of a loop
    visiting an accepting state. Using idempotency of $v$ and similar
    arguments as before, it can be shown that such a loop can be found
    only on three iterations of $v$, \ie on $v^3$. So, once again,
    $P$ can perform a simulation modulo, and use two bits of
    information to check whether it is on the first, second or third
    copy of $v$.

    Note that since $A$ is bounded-visit, and
    since $P$ only performs simulations of $A$ on up to three
    iterations of $v$, it is bounded visit as well. 
\end{proof}

\paragraph{Construction of the \twoPA} We now have all the ingredients
to construct a \twoPA whose language is non-empty if and only if $\Tt$ admits a
critical pattern. More precisely, we show the following:

\begin{lem}\label{lem:critpa}
    The language $\text{Crit}_\Tt$ is recognisable by a \twoPA $C$ of polynomial
    size (in $\Tt$). 
\end{lem}

\begin{proof}
    We do not give a formal construction of $C$, which would be
    unnecessarily too technical. Instead, we explain how $C$ can check
    the six conditions $(1)-(6)$ of the definition of a critical pattern when
    reading valid encodings. The dimension of $C$ is $2$ (\ie its
    vectors are in $\mathbb{N}^2$ or equivalently, it has two
    incrementing counters). We say that the counters are \emph{unchanged} by
    a transition if the vector on that transition is $(0,0)$.

    First $C$ checks, by doing
    a left-to-right pass on the whole input $w$, whether $w$ is a valid
    encoding, \ie whether $w\in {\vdash}((\Sigma\times
    Q_P^2)^*\#)^2((\Sigma\times Q_P)^*\#)^4{\dashv}$. If this is the case,
    then $C$ comes back to the beginning of $u$ and proceed by
    checking other properties, otherwise it rejects. Note that this
    first verification only needs a constant number of states. During
    this phase, counters are unchanged.

    Let us now assume that $C$ has validated the type of its input $w$, which is then of
    the form $w = {\vdash}(u \otimes a_1 \otimes a_2)\# (v \otimes b_1\otimes b_2)\#
    w_1\# z_1\# w_2\# z_2\#{\dashv}$. Let $u_i = u\otimes a_i$ and $v_i =
    v\otimes v_i$. Note that $C$ can navigate between each
    $\#$-free factors of $w$ by just keeping in its state
    whether it is in the first, second, up to sixth factor. This
    information can be updated when the separator $\#$ is met and is
    polynomial. The second observation is that 
    condition $(1)$ ($u_1$ and $u_2$ projects on the same
    $\Sigma$-word, as well as $v_1$ and $v_2$) is necessarily satisfied by definition of the
    encoding.

    Let us explain how $C$ checks condition $(2)$ about
    idempotency of $v_1$ and $v_2$. We rely on
    the construction of Lemma~\ref{lem:idempotentregular} applied to
    the \twoDBT $\widetilde{\Tt}$ where outputs are ignored
    (therefore, it is a deterministic two-way automaton). If this
    second phase accepts, $C$ proceeds to checking conditions
    $(3)-(6)$. Otherwise it rejects. Again, during this verification,
    counters are unchanged.

    We now turn to conditions $(3)-(6)$. The \twoPA $C$ first
    constructs a finite run $\rho_1$ on $u_1v_1w_1$ which does not end
    in $v_1$. Since $\rho_1$ is existentially quantified in the
    definition of critical patterns, this is where non-determinism of
    $C$ is
    needed. To do so, $C$ simulates $\widetilde{\Tt}$ on
    $u_1v_1w_1$. More precisely, it reads $\vdash (u \otimes a_1\otimes
    a_2)\# (v\otimes b_1\otimes b_2)\# w_1\#$ but ignores the third
    components $a_2$ and $b_2$ as well as the $\#$ symbols. During
    this simulation, outputs of $\Tt$ are also ignored. If $\Tt$ wants
    to cross the left boundary of $w_1$, then $C$ rejects, otherwise
    condition $(4)$ is not satisfied. If $\Tt$ is in $v_1$ but
    produces a non-empty output, then $C$ rejects as well, ensuring
    condition $(5)$. Non-deterministically, $C$ decides to stop the
    simulation, but only if $\Tt$ is not in $v_1$ (otherwise, this would violate
    condition $(4)$). If it does not stop the simulation, then $C$
    will never accept its input. Stopping the simulation means that
    $C$ has constructed a run $\rho_1$ which satisfies conditions
    $(4)$ and $(5)$ (for $i=1$). We call this simulation phase $S_1$. To construct
    a run $\rho_2$ on $u_2v_2w_2$, $C$ proceeds similarly as in
    $S_1$. We call this simulation $S_2$. Note that both simulation
    requires only a polynomial number of states.

    It remains to explain $(i)$ how to check condition $(6)$ and $(ii)$ how to
    check condition $(3)$, \ie whether $\rho_i$ can be extended into accepting
    runs over $u_iv_iw_iz_i^\omega$, for $i=1,2$.

    The two counters are used to check condition $(6)$, \ie to check the existence of a mismatch
    between the outputs of $\rho_1$ and $\rho_2$. To do so, the
    simulation $S_i$, $i=1,2$, is again divided into two modes: in the first
    mode, whenever $\Tt$ produces some output word $\alpha$, the
    $i$th counter of $C$ is incremented by $|\alpha|$, using the
    vector $(|\alpha|,0)$ for $i=1$, and $(0,|\alpha|)$ for $i=2$. So,
    the $i$th counter counts the length of the outputs produced so far
    by $\Tt$ on $u_iv_iw_i$. Non-deterministically, it eventually
    decides to stop
    counting this length, and increments the counter  $c_i$ one last
    time by some $0\leq \ell \leq |\alpha|$. It then keeps in memory
    the $c_i$th letter produced by $\Tt$, say,
    $\tau_i\in\Gamma$. After this non-deterministic guess, which
    corresponds to existentially quantify an output position, the
    $i$th counter is left unchanged during the whole computation of
    $C$. The automaton $C$ will accept only if $c_1=c_2$ and $d_1\neq
    d_2$, ensuring the existence of a mismatch between $\out{\rho_1}$
    and $\out{\rho_2}$. Again, this needs only a polynomial number of
    states: simulation $S_i$ is divided into two modes (before and
    after the non-deterministic guess), so, $C$ only
    needs to remember in which mode it is. 

    To conclude, it remains to show how condition $(3)$ can be
    checked, \ie how $C$ can check that $\rho_i$ can be extended
    into an accepting run over $u_iv_iw_iz_i^\omega$. This is a direct
    application of Lemma~\ref{lem:lasso}. 
\end{proof}

Finally, Lemma~\ref{lem:decpat} is a direct consequence of Lemma~\ref{lem:critpa} and
Proposition~\ref{prop:trivial}. 

\paragraph{Deciding continuity for regular functions} We prove the
following theorem:

\begin{thm}\label{thm:main2}
    Deciding whether a regular function given as a deterministic
    two-way transducer with prophetic B\"uchi look-ahead is
    continuous is \textsc{PSpace}-c. 
\end{thm}

\begin{proof}
    The lower bound has been established in
    Theorem~\ref{thm:main1}. For the upper-bound, let $\Tt$ be a
    \twoDFTpla defining a regular function $f$, with set of look-ahead
    states $Q_P$. According to
    Proposition~\ref{prop:criticalpatterntwoway}, one needs to check
    whether $\widetilde{\Tt}$ admits a critical pattern
    $u_1,v_1,w_1,z_1,u_2,v_2,w_2,z_2$ such that
    $\pi_\Sigma(u_1v_1^\omega)\in\dom{f}$. We have seen in the proof
    of Theorem~\ref{thm:main1} how to construct a \twoPA $P$ of polynomial
    size which accepts
    the set of (encodings of) critical patterns. Additionally here,
    $P$ needs to check whether
    $\pi_\Sigma(u_1v_1^\omega)\in\dom{f}$. The difficulty lies in that 
    $\pi_\Sigma(u_1v_1^\omega)\in\dom{f}$ is not equivalent to 
    $u_1v_1^\omega\in\dom{\tilde{f}}$, because $u_1v_1^\omega$ may not be a good annotated word (see
    Section~\ref{subsubsec:la}). So, it is not correct to check whether
    $u_1v_1^\omega\in\dom{\tilde{f}}$.

   To overcome this difficulty, we modify the notion of critical
   pattern into \emph{strong} critical pattern, which also includes
   some good annotation of $\pi_\Sigma(u_1v_1^\omega)$. A
    strong critical pattern is a tuple of words
    $(u'_1,v'_1,w_1,z_1,u_2,v_2,w_2,z_2)$ such that $u'_1,v'_1\in
    (\Sigma\times Q_P\times Q_P)^*$,
    $(\pi_{1,2}(u'_1),\pi_{1,2}(v'_1),w_1,z_1,u_2,v_2,w_2,z_2)$ is a critical
    pattern and $\pi_{1,3}(u'_1(v'_1)^\omega)\in\dom{\tilde{f}}$, where
    $\pi_{1,2}$ is a morphism which only keeps the first and second
    alphabet components while $\pi_{1,3}$
    only keeps the first and third components. We show that $f$ is
    continuous if and only if $\Tt$ admits a strong critical pattern. If $\Tt$
    admits a strong critical pattern
    $(u'_1,v'_1,w_1,z_1,u_2,v_2,w_2,z_2)$, then for $u_1 =
    \pi_{1,2}(u'_1)$ and $v_1 = \pi_{1,2}(v'_1)$, we have that
    $(u_1,v_1,w_1,z_1,u_2,v_2,w_2,z_2)$ is a critical
    pattern. Moreover,
    $\pi_{1,3}(u'_1(v'_1)^\omega)\in\dom{\tilde{f}}$, which
    implies that $\pi_\Sigma(u'_1(v'_1)^\omega) =
    \pi_\Sigma(u_1v_1^\omega)\in\dom{f}$. Hence $f$ is continuous.

    Conversely, if $f$ is continuous, $\widetilde{\Tt}$ admits a
    critical pattern $pat = (u_1, v_1, w_1, z_1, u_2, v_2$, $w_2, z_2)$ such that
    $\pi_\Sigma(u_1v_1^\omega)\in\dom{f}$. Since the look-ahead
    automaton is prophetic, its accepting run on $\pi_\Sigma(u_1v_1^\omega)$  is of the form
    $r = r_1r_2^\omega$ such
    that $r_1$ is a run on $\pi_\Sigma(u_1v_1^{k_1})$ for some $k_1\geq 0$ and $r_2$
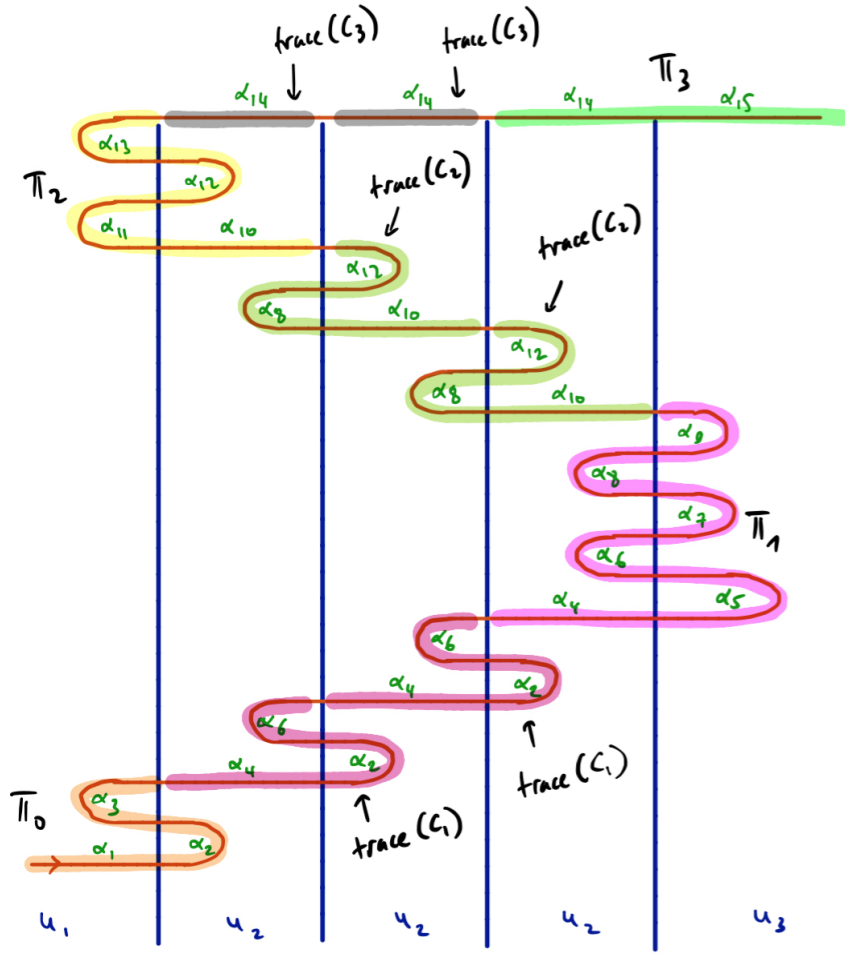
    is a run on $\pi_\Sigma(v_1^{k_2})$ for some $k_2\geq 1$. We then
    let $u'_1 = (u_1v_1^{k_1})\otimes r_1$ and $v'_1 =
    (v_1^{k_2})\otimes r_2$. We also let $u'_2 = u_2v_2^{k_1}$ and
    $v'_2 = v_2^{k_2}$.  We claim that
    $(u'_1,v'_1,w_1,z_1,u'_2,v'_2,w_2,z_2)$ is a strong critical
    pattern. Clearly, $\pi_{1,3}(u'_1(v'_1)^\omega)\in
    \dom{\tilde{f}}$, because the annotation of
    $\pi_{1,3}(u'_1(v'_1)^\omega)$ is $r_1r_2^\omega$ which is an
    accepting run of the look-ahead automaton. It remains to show that 
    $pat' = (\pi_{1,2}(u'_1) = u_1v_1^{k_1},\pi_{1,2}(v'_1)=v_1^{k_2},w_1,z_1,u'_2=u_2v_2^{k_1},v'_2=v_2^{k_2},w_2,z_2)$ is a
    critical pattern for $\widetilde{\Tt}$. It is immediate as  $v_1$ and
    $v_2$ are idempotent and non-producing, so, iterating them does
    not change the existence of accepting runs whose outputs mismatch.

    We have established that $f$ is continuous if and only if $\widetilde{\Tt}$
    admits a strong critical pattern. As shown in 
     Lemma~\ref{lem:critpa}, the set of (encodings of) critical
    patterns is recognisable by a bounded-visit \twoPA of polynomial size. We
    need here to slightly modify the \twoPA constructed in Lemma~\ref{lem:critpa} so that it
    also checks whether $\pi_{1,3}(u'_1(v'_1)^\omega)$ is in the
    domain of $\Tt$. Based on Lemma~\ref{lem:lasso}, we can construct
    a bounded-visit two-way automaton of polynomial size checking the
    latter property. When this automaton has successfully checked the
    latter property, it is then sequentially composed with a
    \twoPA as in Lemma~\ref{lem:critpa} checking whether the input
    (where the third components of the alphabets are ignored) forms a
    critical pattern. The resulting composition is a bounded-visit
    \twoPA of polynomial size, whose emptiness can be checked in
    \textsc{PSpace}, as shown in~\cite{DBLP:conf/fsttcs/FiliotGM19}.
\end{proof}

\section{Conclusion}

In this paper, we have studied two notions of computability for
rational and regular functions, shown their
correspondences to continuity notions which we proved to be
decidable in \textsc{NLogSpace} and \textsc{PSpace} respectively. The notion of uniform computability asks for the existence
of a modulus of continuity, 
which tells how far one has to go in the input to produce a certain amount of
output. It would be interesting to give a tight upper bound on modulus
of continuity for regular functions, and we conjecture that it is always a linear (affine)
function in that case.

Another interesting direction is to find a transducer
model which captures exactly the computable, and uniformly
computable, rational and regular functions. For rational functions,
the deterministic (one-way) transducers are not sufficient, already
for uniform computability, as
witnessed by the rational function which maps any word of
the form $a^nb^\omega$ to itself, and any word $a^nc^\omega$ to
$a^{2n}c^\omega$. For regular
functions, we conjecture that \twoDFT characterise the computable
ones, but we have not been able to show it yet.

Finally, much of our work deals with reg-preserving functions in
general. An interesting line of research would be to investigate
continuity and uniform continuity for different classes of functions
which have this property. One natural candidate is the class of
\emph{polyregular functions} introduced in \cite{DBLP:journals/corr/abs-1810-08760} which enjoy
several different characterisations and many nice properties,
including being effectively reg-preserving. This means that continuity
and computability also coincide, however deciding continuity seems
challenging.

\bibliographystyle{alphaurl}
\bibliography{papers}

\newpage
\appendix
\input{appendix}

\end{document}

%% file: appendix.tex
\section{Proofs from Section~\ref{sec:prelim}}
\label{app:equitrans}
\begin{proof}[Proof of Theorem~\ref{thm:equivtrans}]
	In~\cite{lics12}, authors show that a function is regular iff it is definable by $\twoDFTla$.
    We show that a function is $\twoDFTla$ definable iff it is $\twoDFTpla$ definable.
    Recall that $\twoDFTla$ is defined as $(\Tt, A, B)$ where $\Tt$ is a two-way deterministic transducer with Muller acceptance, $A$ is a regular look-ahead automaton, given by a Muller automaton and $B$ is a regular look-behind given by a finite state automaton.
	
	Given a \twoDFTpla $(\Aa, P)$, where $\Aa = (Q, \Sigma, \Gamma, \delta_{\Aa}, q_0)$ and $P=(Q_P, \Sigma, \delta_P, Q_0, F)$, we  construct a \twoDFTla $(\Tt, A, B)$ as follows: For every state $p \in Q_P$, we construct an equivalent Muller automaton $A_p$ with initial state $s_p$ s.t. $\Lang(P, p) = \Lang(A_p)$.
	The Muller look-ahead automaton $A$ used in the  \twoDFTla $(\Tt, A, B)$ is the  disjoint union of the Muller automata $A_p$ for all $p \in Q_P$.
	$\Tt= (Q, \Sigma, \Gamma,  \delta_{\Tt}, q_0, 2^Q\backslash  \emptyset)$, and $\delta_{\Tt}$ is obtained by 
	 modifying the transition function $\delta_{\Aa}(q, a, p) = (q', \gamma, d)$ as $\delta_{\Tt}(q, a, s_p) = (q', \gamma, d)$.
	Since the language accepted by two distinct states of a prophetic automaton are disjoint, 
	the language accepted by $A_p$ and $A_p'$ are disjoint for $p \neq p'$. 
	The look-behind automaton $B$ accepts all of $\Sigma^*$. It is easy to see that 
	   $(\Tt, A, B)$ is deterministic : on any position $i$ of the input word $a_1a_2 \dots$, and any state $q \in Q$, there is a unique 
	   $A_p$  accepting $a_{i+1}a_{i+2} \dots$. The domain of  $(\Tt, A, B)$ is the same as that of 
	   $(\Aa, P)$, since the accepting states of $A$ are the union of the accepting states of all the $A_p, p \in Q_P$. 
	   Since all the transitions in
	   $\delta_{\Tt}$ have the same outputs as in $\delta_{\Aa}$ for each $(q,a,p)$, the function computed by  $(\Aa, P)$ is the same as that computed by $(\Tt, A, B)$.

	For the other direction, given a \twoDFTla,  it is easy to remove the look-behind  by a product construction \cite{DBLP:journals/corr/abs-1802-02094}. Assume we start 
	with such a modified \twoDFTla $(\Tt, A)$ with no look-behind.   Let $\Tt=(Q_{\Tt}, \Sigma, \Gamma, \delta_{\Tt}, s_{\Tt}, \mathcal{F}_{\Tt})$ and $A=(Q_A,\Sigma, \delta_A, s_A, 
	\mathcal{F}_A)$ where $\mathcal{F}_{\Tt}, \mathcal{F}_A$ respectively are the Muller sets corresponding to  
	 $\Tt$ and $A$. 
		We describe how to obtain a 
	corresponding \twoDFTpla $(\Aa, P)$ with $P$, a prophetic \buchi{} look-ahead automaton. 
	$\Aa$ is described as $(Q_{\Tt}, \Sigma, \Gamma, \delta, p_0)$. 
		The prophetic look-ahead automaton is described as follows.   
		Corresponding to each state $q \in Q_A$, let  $P_q$ be a prophetic automaton such that  $\Lang(A, q) = \Lang(P_q)$
		(this is possible since prophetic automata capture $\omega$-regular languages \cite{DBLP:conf/birthday/CartonPP08}). 
				 Since $\Aa$ has no accepting condition, we also 
		consider a prophetic look-ahead automaton to capture 		$\dom{\Tt}$. The Muller acceptance of $\Tt$ can be translated to a 
		B\"uchi acceptance condition, and let $P_{\dom{\Tt}}$ represent the prophetic automaton 
		  such that $\Lang(P_{\dom{\Tt}})=\dom{\Tt}$.
		  Thanks to the fact that prophetic automata are closed under synchronized product, the prophetic automaton $P$ we need, is the product of $P_q$ for all $q \in Q_A$ and $P_{\dom{\Tt}}$.  
		  Assuming an enumeration $q_1, \dots, q_n$ of $Q_A$,  
		  the states of $P$ are $|Q_A|+1$ tuples where the first $|Q_A|$ entries correspond to states of $P_{q_1}, \dots, P_{q_n}$, and 
		  the last entry is a state of $P_{\dom{\Tt}}$. 
		  		   Using this prophetic automaton $P$ and transitions $\delta_{\Tt}$, 
		  we obtain the transitions $\delta$ of $\Aa$ as follows. 
		  Consider a transition $\delta_{\Tt}(p, a, q_i) = (q', \gamma, d)$. Correspondingly in $\Aa$, we have 
		  $\delta(p,a,\kappa)=(q', \gamma, d)$ where $\kappa$ is a $|Q_A|+1$ tuple of states 
		  such that the $i$th entry of $\kappa$ is an initial state of $P_{q_i}$.
		  From the initial state $s_{\Tt}$, on reading $\leftend$,  
		    if we have $\delta_{\Tt}(p_0, \leftend, q_i) = (q', \gamma, d)$, then  $\delta(p_0,\leftend,\kappa)=(q', \gamma, d)$ such that the $i$th entry of $\kappa$ is an initial state of $P_{q_i}$ and the last   
		  entry of $\kappa$ is an initial state of $P_{\dom{\Tt}}$.
		  
	To see 	why $(\Aa, P)$ is deterministic.  
	For each state $q \in Q_{\Tt}$, for each position $i$ in the input word $a_1a_2 \dots a_i \dots$, 
	there is a unique state $p \in Q_A$  such that $a_{i+1}a_{i+2}\dots$ is accepted by $A$. By our construction, the language accepted from 
	each $p \in Q_A$ is captured by the prophetic automaton $P_p$; by the property of the prophetic automaton $P$, we know that 
	for any $q_1, q_2 \in Q_A$ with $q_1 \neq q_2$, $L(P_{q_1}) \cap L(P_{q_2}) = \emptyset$. Thus, in $(\Aa,P)$, 
	for each state $q$ of $\Aa$, there is a unique state $p \in P$ such that the suffix is accepted 
	by $L(P)$; further, from the initial state of $\Aa$, from $\leftend$, there is a unique state $p \in P$ which accepts 
	$\dom{(\Tt,A)}$. 	Hence $\dom{(\Aa,P)}$ is exactly same as $\dom{(\Tt,A)}$. 
	For each $\delta_{\Tt}(q,a,q_i)=(q', \gamma, d)$, we have the transition 
	$\delta(q,a,\kappa)=(q', \gamma, d)$ with the $i$th entry of $\kappa$ equal to the initial state of $P_{q_i}$, 
	which preserves the outputs on checking that the suffix of the input from the present position is in $L(P_{q_i})$. Notice that 
	the entries $j \neq i$ of $\kappa$ are decided uniquely, since 
	there is a unique state in each $P_q$ from where each word has an accepting run.  
	 Hence, $(\Aa,P)$ and $(\Tt,A)$ capture the same function. 
\end{proof}

\begin{rem}
	Example function $g$ given in Section~\ref{sec:prelim}, after Theorem~\ref{thm:equivtrans} cannot be realised without look-ahead, not even if we allow non-determinism. 
\end{rem}